\documentclass[12pt]{iopart}
\usepackage{graphicx}

\begin{document}
\def\etal{\emph{et al.}}
\def\Ek{$E(\mathbf{k})$}
\def\EF{$E_{\mathrm{F}}$}
\def\ED{$E_{\mathrm{D}}$}
\def\Ei{$E_{i}$}
\def\DED{$\Delta E_{\mathrm{D}}$}
\def\TC{$T_{\mathrm{C}}$}
\def\CaC{$C_{\mathrm{6}}$}
\def\C6{$C_{\mathrm{6}}$}
\def\K2by{$K-(2 \times 2)$}
\def\rt3{$\sqrt{3}\times\sqrt{3}$-R30}
\def\sixr3{$6\sqrt{3}\times 6\sqrt{3}$}
\def\k{$\mathbf{k}$}
\def\kpar{$k_{\parallel}$}
\def\kperp{$k_{\perp}$}
\def\ky{$k_{\mathrm{y}}$}
\def\kx{$k_{\mathrm{x}}$}
\def\kspace{$\mathbf{k}$-space}
\def\kKpoint{\kpar$=1.703~$\AA$^{-1}$}
\def\A1{\AA$^{{-1}}$}
\def\vF{$v_{\mathrm{F}}$}
\def\me{$m_{\mathrm{e}}$}
\def\red{\textcolor{red}}
\def\blue{\textcolor{blue}}
\def\c60{C$_{60}$}
\def\gk{$\mathrm{\Gamma}$K}
\def\g{$\mathrm{\Gamma}$}
\def\g1{$\gamma_{1}$}
\def\twoby{$2\times 2$}
\def\kc{KC$_{8}$}
\def\ims{Im$\mathrm{\Sigma}$}
\def\res{Re$\mathrm{\Sigma}$}
\def\imsw{Im$\mathrm{\Sigma}(\omega)$}
\def\resw{Re$\mathrm{\Sigma}(\omega)$}
\def\akw{$A(\mathbf{k},\omega)$}
\def\skw{${\mathrm{\Sigma}(\mathrm{\mathbf{k}},\omega)}$}
\def\imskw{${\mathrm{Im\Sigma}(\mathrm{\mathbf{k}},\omega)}$}
\def\reskw{${\mathrm{Re\Sigma}(\mathrm{\mathbf{k}},\omega)}$}
\def\n56{$n=5.6\times 10^{13}$ cm$^{-2}$}
\def\sw{${\mathrm{\Sigma}(\omega)}$}
\def\wb{$\omega_{\mathrm{b}}(\mathbf{k})$}
\def\wbk{$\omega_{\mathrm{b}}(\mathrm{\mathbf{k}})$}
\def\wdx{$\omega_{\mathrm{D}}$}
\def\wph{$\omega_{\mathrm{ph}}$}
\def\vph{$v_{\mathrm{ph}}$}
\def\vf{$v_{\mathrm{F}}$}
\def\vd{$v_{\mathrm{D}}$}
\def\pz{$p_{z}$}
\def\pistar{$\pi^{*}$}
\def\zeroth{$0^{th}$}

\def\mr{$m_{\mathrm{r}}$}
\def\me{$m_{\mathrm{e}}$}
\def\leedsym{$6\sqrt{3}\times 6\sqrt{3}$}
\def\gm{$\mathrm{\Gamma}$M}
\def\eh{$e$-$h$}
\def\ee{$e$-$e$}
\def\eph{\emph{e}-\emph{ph}}
\def\epl{\emph{e}-\emph{pl}}
\def\hwph{$\hbar\omega_{\mathrm{ph}}$}
\def\wpl{$\omega_{\mathrm{pl}}$}
\def\thomasAddr{Institut f\"{u}r Physik der Kondensierten Materie, Lehrstuhl f\"{u}r Technische Physik, Universit\"{a}t Erlangen-N\"{u}rnberg, 
Erwin-Rommel-Stra\ss e 1, D-91058 Erlangen, Germany}
\def\af{$\alpha^{2}F(\omega)$}

\def\mathimskw{\mathrm{Im\Sigma}(\mathrm{\mathbf{k}},\omega)}
\def\mathreskw{\mathrm{Re\Sigma}(\mathrm{\mathbf{k}},\omega)}
\def\mathakw{A(\mathrm{\mathbf{k},\omega})}

\def\figAtomsText{Atomic arrangement in (left) monolayer and (right) bilayer graphene. The inset shows the unit cell with two
equivalent atoms.}
\def\figAtoms{\begin{figure}\center{\includegraphics[width=6in]{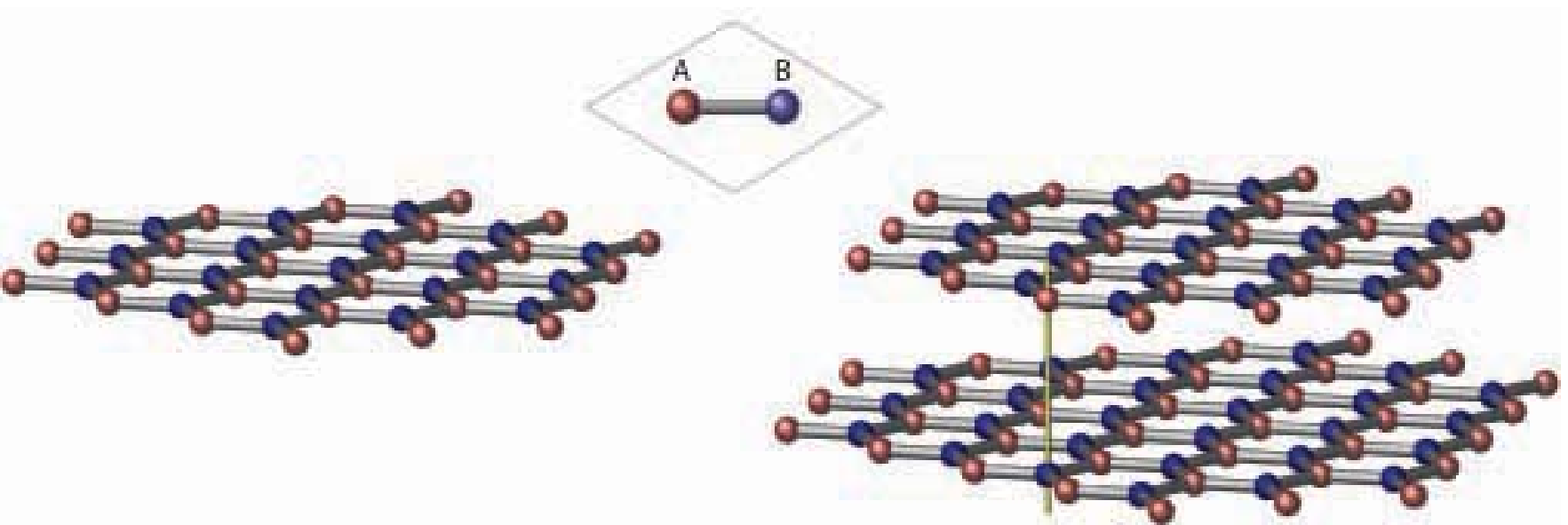}}\caption{\label{f:figAtoms}\figAtomsText}\end{figure}}

\def\figBandtext{Theoretical tight-binding band structure for graphene, based on third NN parameters due to
Reich\cite{reich2002}.}
\def\figBand3d{\begin{figure}\center{\includegraphics[width=4in]{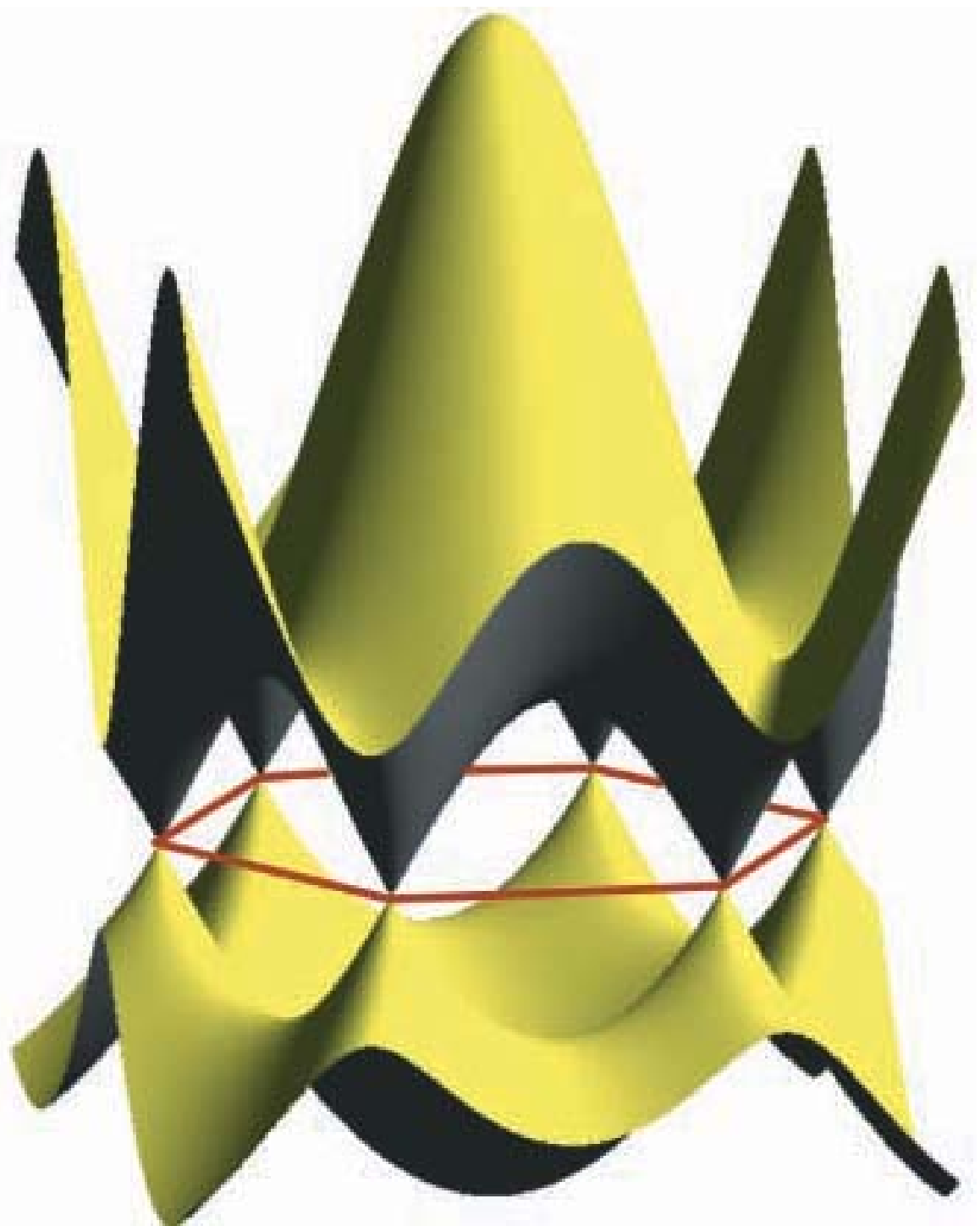}}\caption{\label{f:figBand3d}\figBandtext}\end{figure}}

\def\figExptMLtext{Experimental Fermi surfaces (left) and band structures (middle, right) for (a) as-prepared monolayer graphene 
and (b) graphene dosed with K atoms.  The middle and rightn panels are taken along orthogonal directions through the K point as 
indicated. Adapted from Ref. \cite{bostwick2007}.   The doping levels in electrons per cm$^{2}$ are indicated. The phonon kinks at
$\sim 200$ meV are indicated by arrows.}
\def\figExptML{\begin{figure}\center{\includegraphics[width=6in]{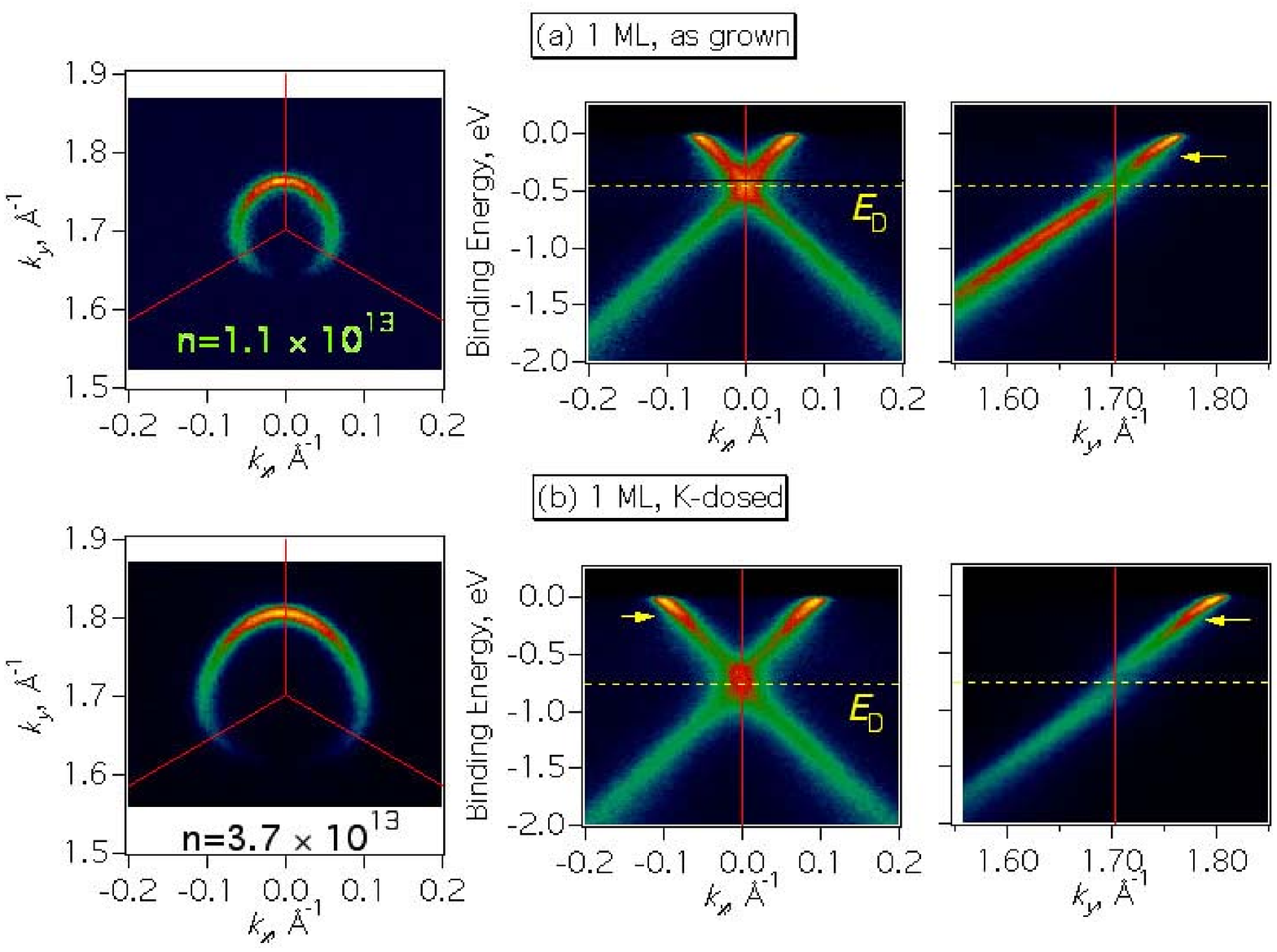}}\caption{\label{f:figExptML}\figExptMLtext}\end{figure}}

\def\figNoGaptext{(a) Bandstructure for as-prepared graphene.  The dashed lines are a projection of the $\pi$ bands and highlight
the fact that this projection does not pass through the $\pi*$ bands above \ED. (b) The momentum-integrated density of states,
derived by integrated the bandstructure in (a).  No dip in the density of states is observed at the Dirac crossing point.  (c) The
individual energy distribution curves for the bands in (a).  The center EDC that includes the Dirac crossing shows no resolved
splitting, which would be expected for a gap.}
\def\figNoGap{\begin{figure}\center{\includegraphics[width=5in]{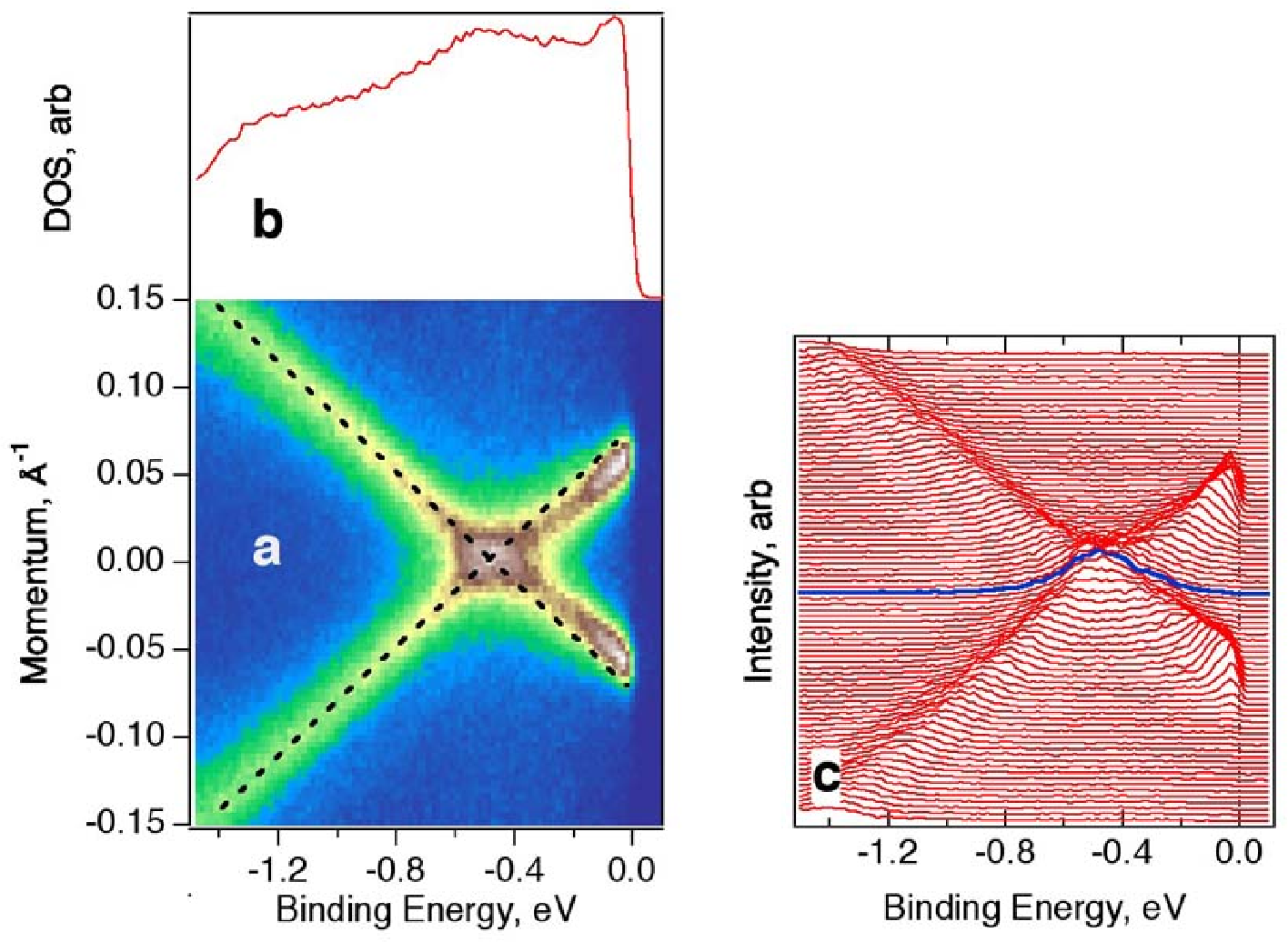}}\caption{\label{f:figNoGap}\figNoGaptext}\end{figure}}

\def\figNoInttext{(a) Polar plot of the intensity of the Fermi contours for monolayer (solid circles) and bilayer graphene (open
circles), obtained by fitting momentum distribution curves taken along radial cuts through the K point of the Brillouin zone.  The
intensity scale is logarithmic.  Also shown are theoretical intensities for asymmetry parameters $\Delta=0.0, 0.1, 0.2$ eV (solid
lines) using Shirley's formalism\cite{shirley1995}.  The solid black data point is an upper limit based on the noise floor indicated
by the central yellow circle.  (b) The ratio of the weakest to strongest emission intensities as a function of asymmetry parameter
$\Delta$.  The noise floor (yellow region) establishes the maximum value of the asymmetry parameter $\sim 55$ meV admitted by our
measured intensity distribution.}
\def\figNoInt{\begin{figure}\center{\includegraphics[width=5in]{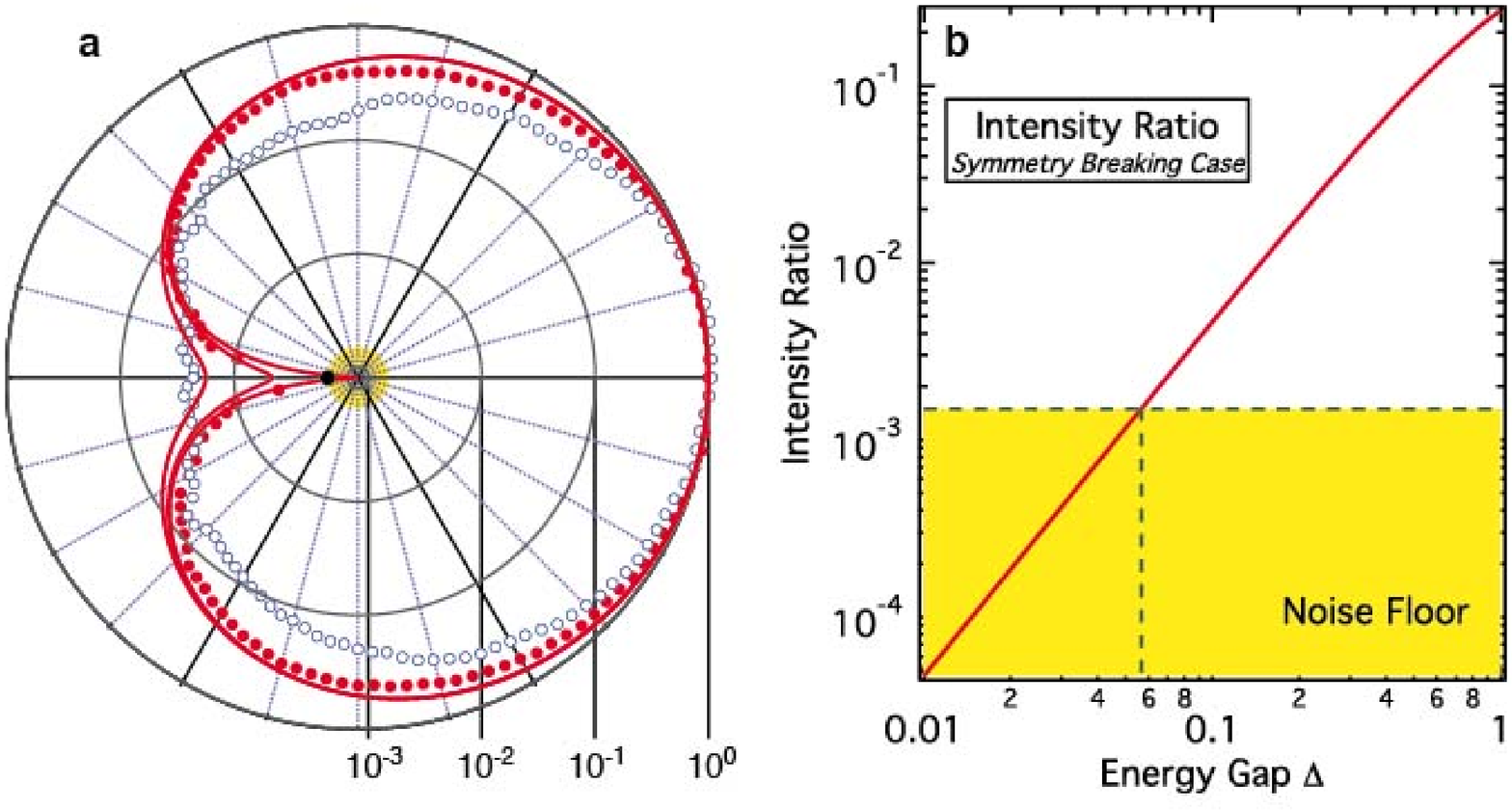}}\caption{\label{f:figNoInt}\figNoInttext}\end{figure}}

\def\figSatText{
Constant energy surface at the Dirac energy \ED\ for (a) ordinary linear intensity scale and (b) highly non-linear intensity scale. 
The weak satellite bands barely visible with a linear scale are highlighted at the right. The remaining background intensity is
attributed to the \sixr3 interface layer. The sample is as-grown graphene, and measurements were at $T\sim25$K.}
\def\figSat{\begin{figure}\center{\includegraphics[width=4in]{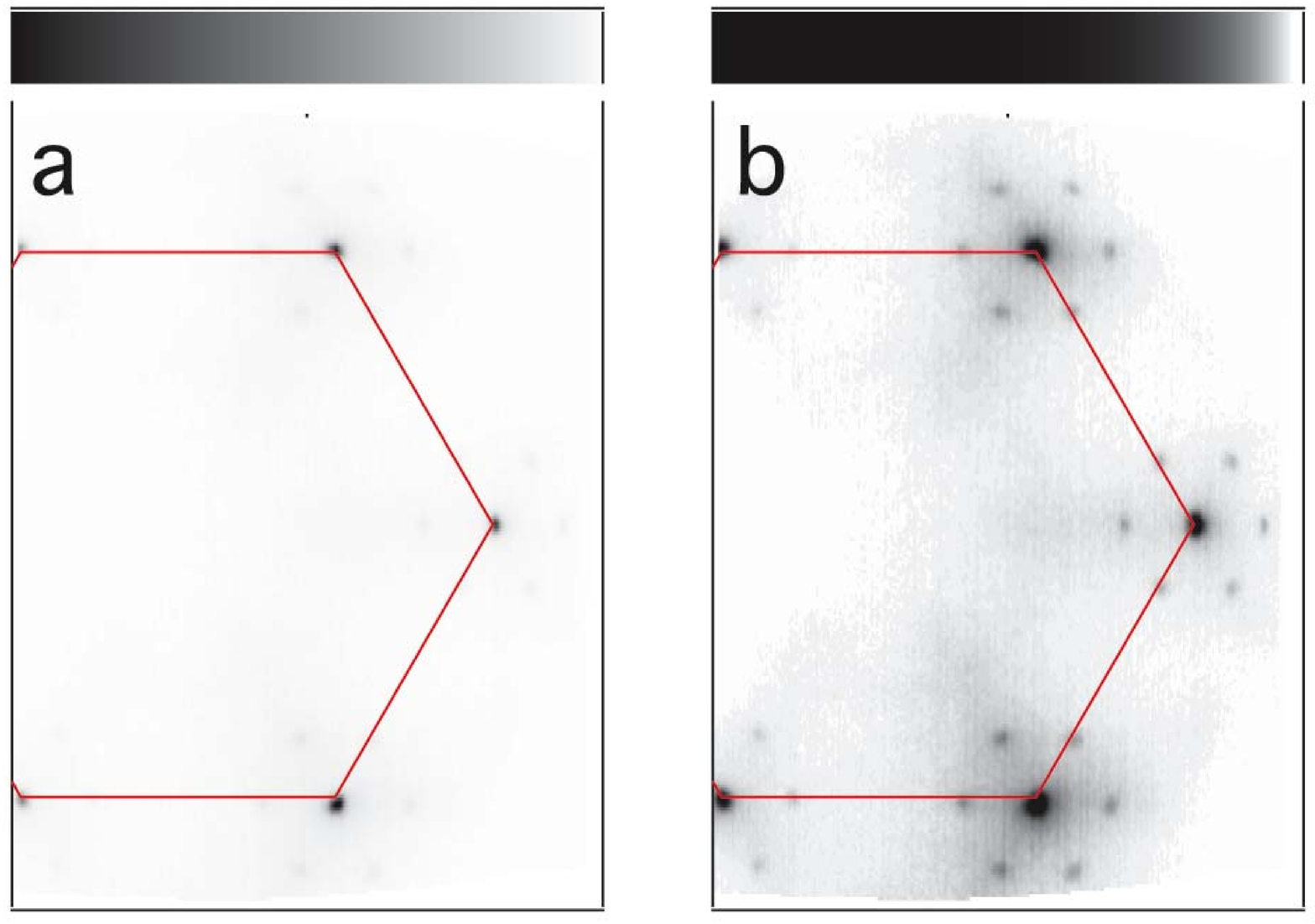}}\caption{\label{f:figSat}\figSatText}\end{figure}}

\def\figSpecHilbertText{Experimentally determined Self-Energy function.  (a) The width of the momentum distribution curves as a
function of energy.  (b) \imskw\ derived from scaling the MDC widths by half the band velocity (black) and smoothing (red).
(c) \reskw\ obtained from the experimental data (black line) and by Hilbert transforming the smoothed \imskw\ (red). The sample is
doped to \n56.}
\def\figSpecHilbert{\begin{figure}\center{\includegraphics[width=3in]{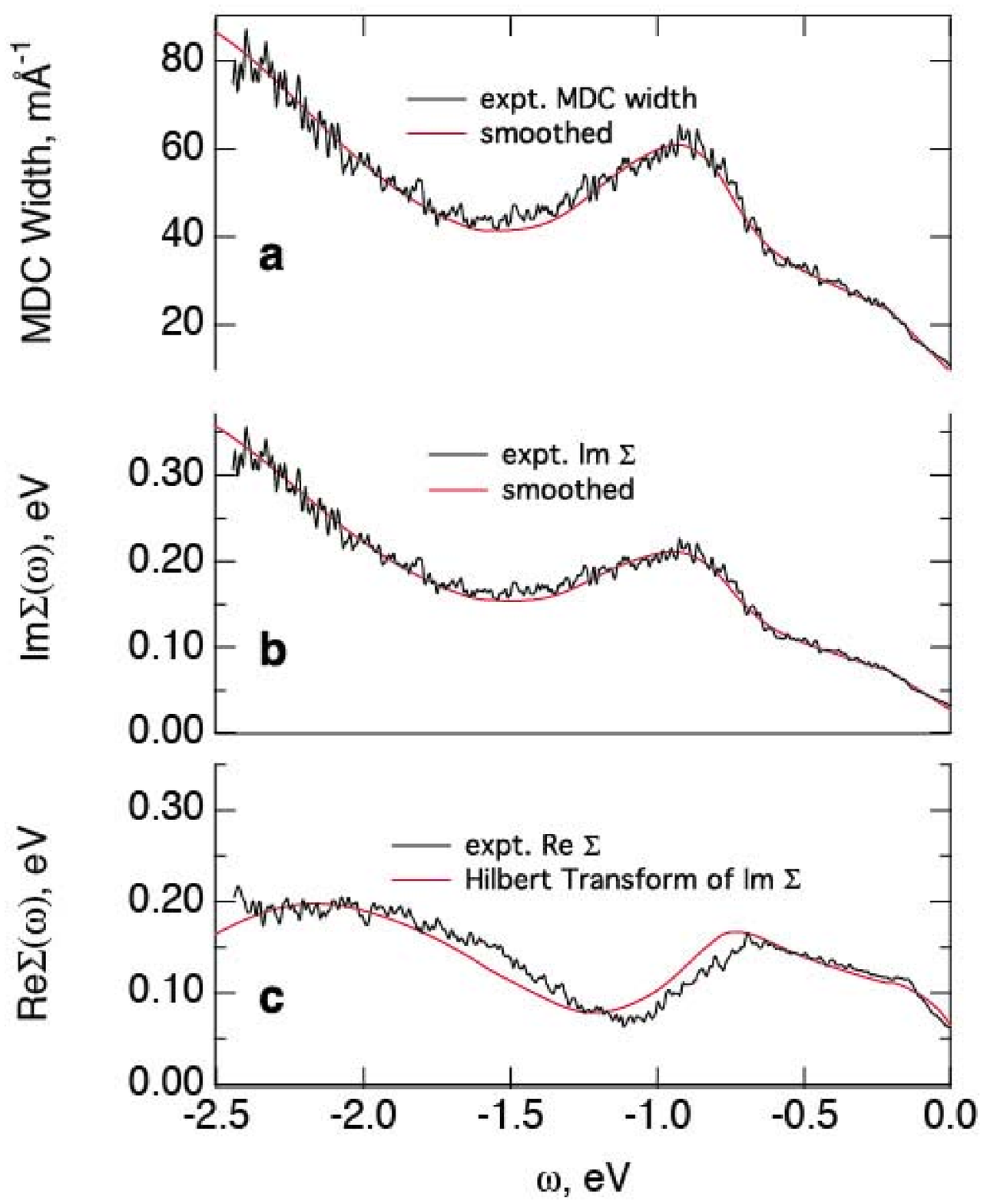}}\caption{\label{f:figSpecHilbert}\figSpecHilbertText}\end{figure}}

\def\figSpecCompText{Spectral function of doped graphene.  (a) The experimentally determined spectral function for graphene doped with 
K atoms (total doping \n56). The solid line is the fitted band position \wb + \reskw, the dotted line is the bare
band \wb.  (b) A model spectral function generated using only the measured \imskw\ and the bare band \wb. Adapted from
Ref.\cite{bostwick2007}.}
\def\figSpecComp{\begin{figure}\center{\includegraphics[width=4in]{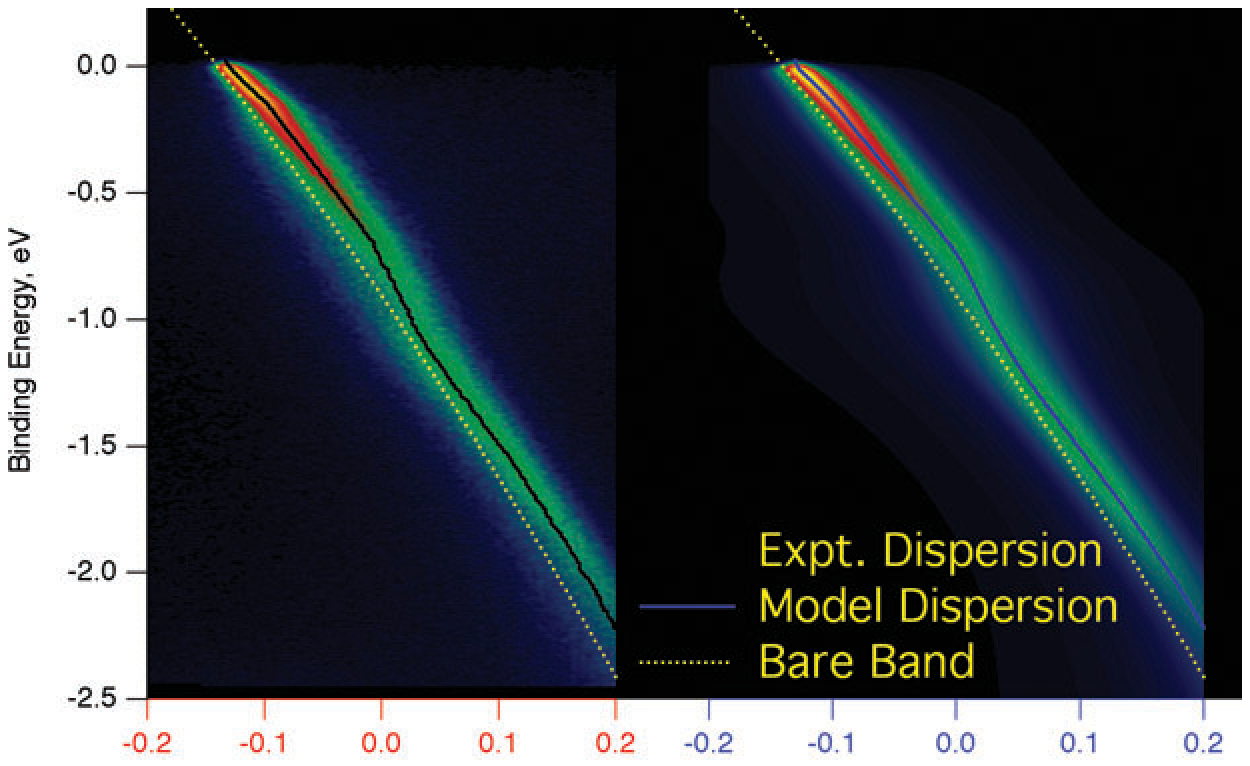}}\caption{\label{f:figSpecComp}\figSpecCompText}\end{figure}}

\def\figSpecTheoryText{Comparison of calculated and measured MDC widths.  (a) measured MDC widths (dots) for the highest-doping
sample (\n56) are compared to the total scattering rate contribution from Bostwick \etal\cite{bostwick2007}(solid).  (b) the
calculated contributions to the scattering rate due to electron-hole pair generation, electron-phonon scattering, and
electron-plasmon scattering\cite{bostwick2007}.  (c-e) experimental MDC widths for $n=$1.2, 3.0, and \n56\ in comparison to the
calculations of Hwang \etal\cite{hwang2006b}.  Adapted from Ref.\ \cite{bostwick2007}.}
\def\figSpecTheory{\begin{figure}\center{\includegraphics[width=4in]{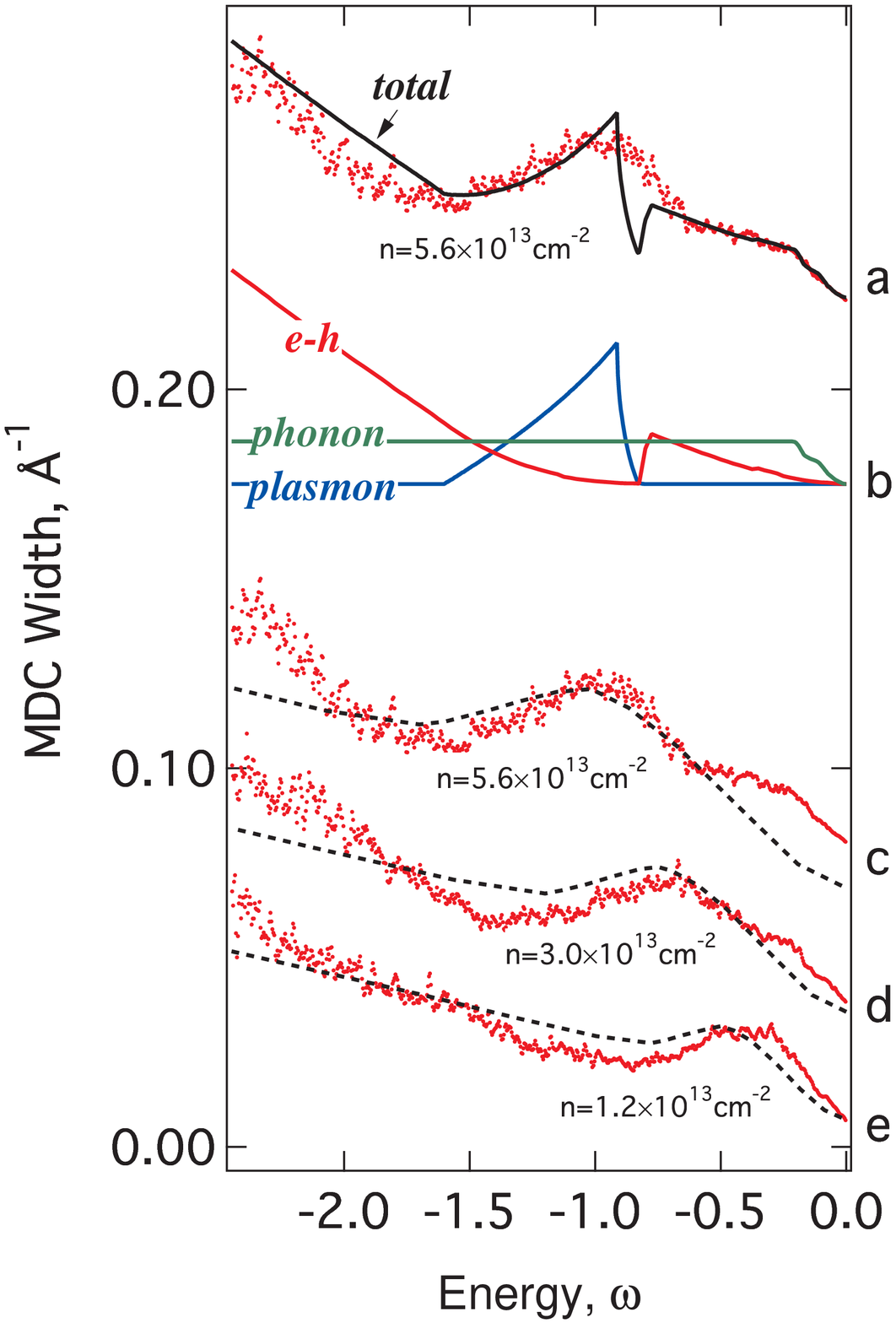}}\caption{\label{f:figSpecTheory}\figSpecTheoryText}\end{figure}}

\def\figProcessText{Possible many-body decay process in $n$-doped graphene.  (a) decay by electron-phonon emission (b) decay be
electron-plasmon emission (c) decay by electron-hole pair generation.}
\def\figProcess{\begin{figure}\center{\includegraphics[width=4in]{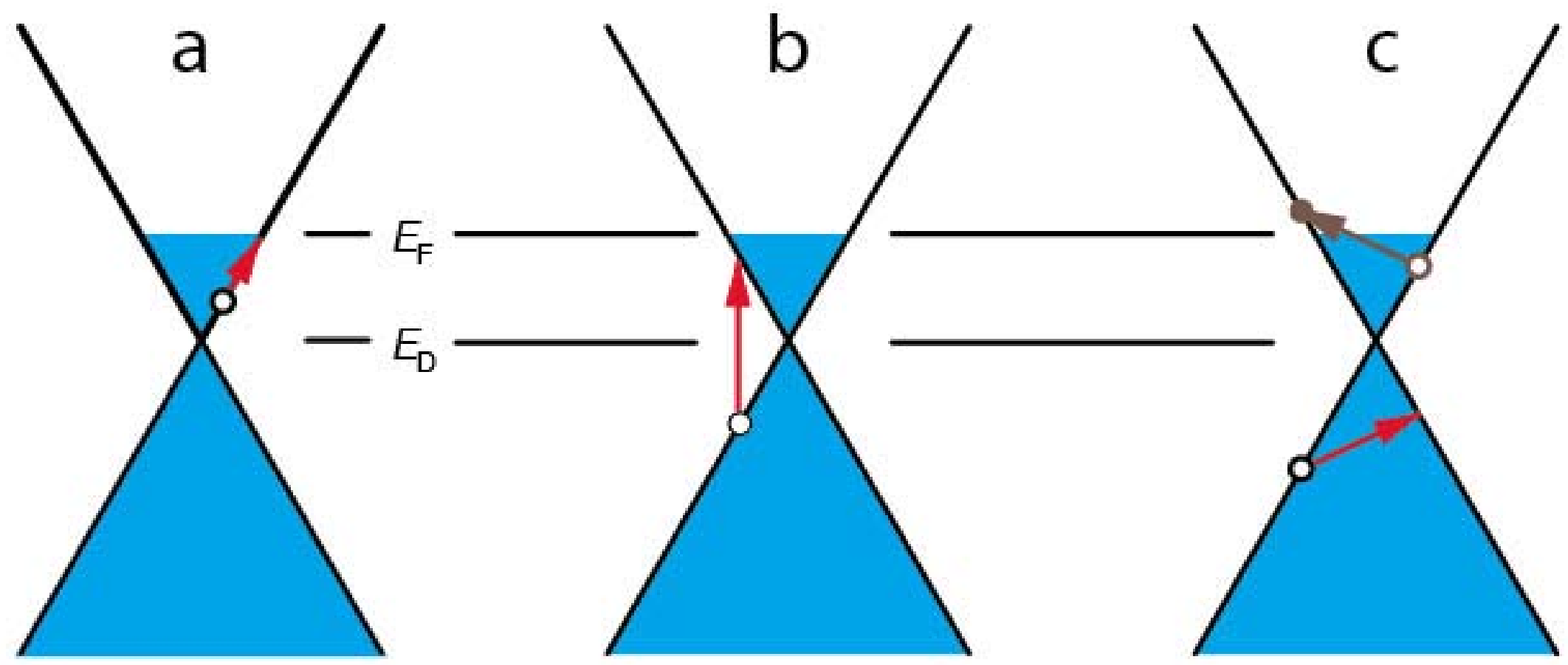}}\caption{\label{f:figProcess}\figProcessText}\end{figure}}

\def\figTheorytext{Calculated band structures for few layer graphene using Eq.\ \ref{e:Hgen}.  (a)-(d) show the bands for 1-4 layers 
graphene, respectively.  The upper panels are for unbiased layers (\Ei=0) while the lower panels were calculated with a 200 meV 
potential difference across the films, assuming a linear field gradient. Calculations were for $\gamma_{1}=0.4$ eV and $v=6.91$ 
eV/\AA$^{-1}$.}
\def\figTheory{\begin{figure}\center{\includegraphics[width=5in]{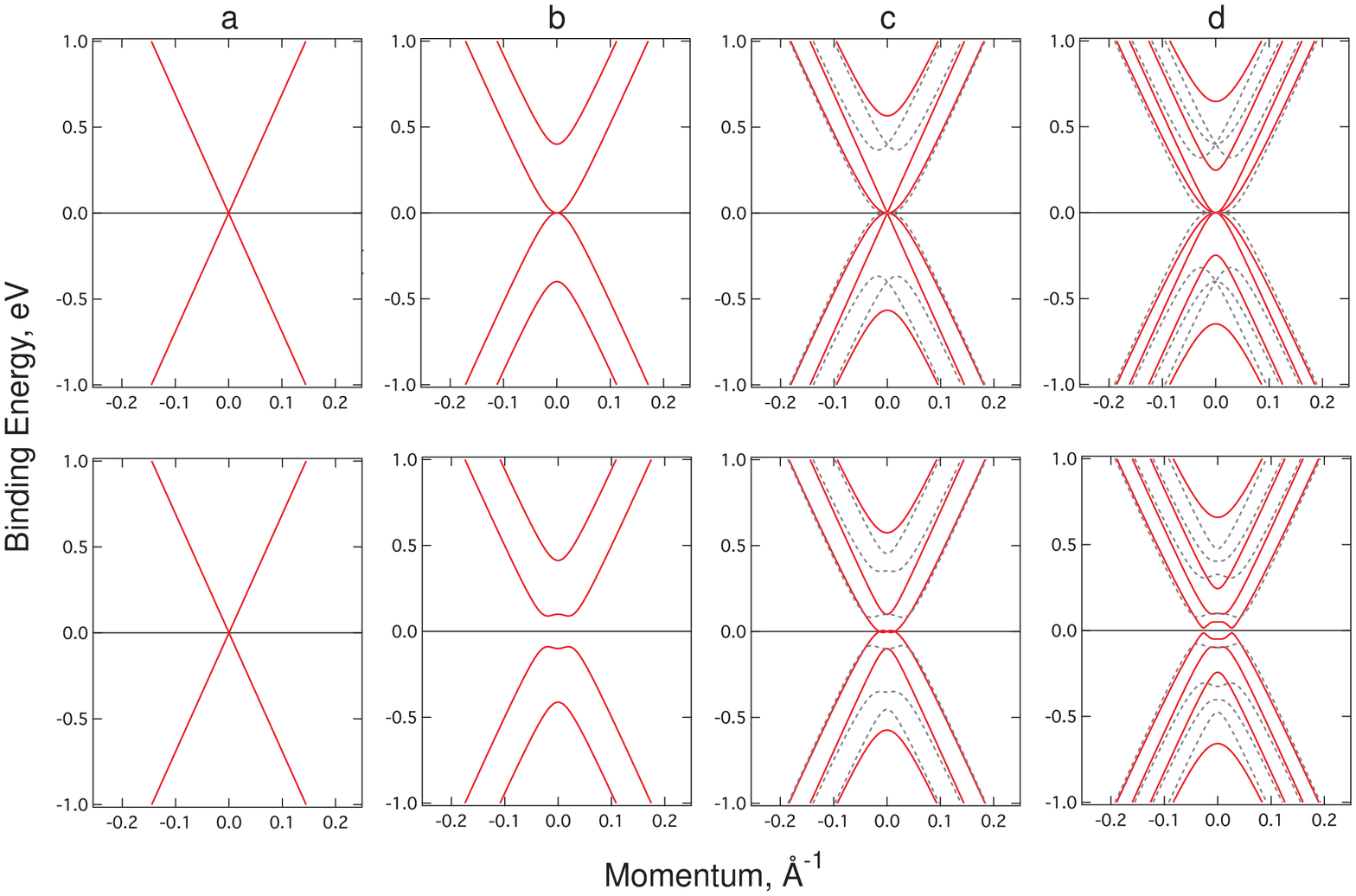}}\caption{\label{f:figTheory}\figTheorytext}\end{figure}}

\def\figTwoLayertext{Gap Control of Bilayer Graphene. (a) An unbiased bilayer has a gapless spectrum, which we could observe for a
doped sample which carefully balanced the field across the film.  (b) For a bilayer with a field gradient, an energy gap is opened 
between $\pi$ and $\pi^{*}$ states.}
\def\figTwoLayer{\begin{figure}\center{\includegraphics[width=5in]{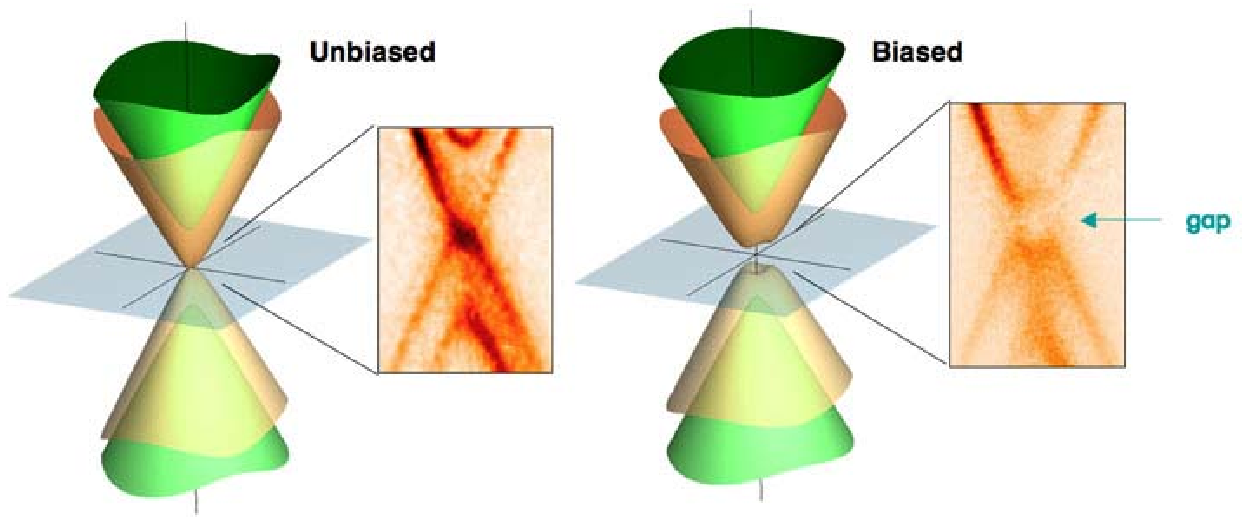}}\caption{\label{f:figTwoLayer}\figTwoLayertext}\end{figure}}

\def\figFLGText{Band structure of graphene films of thickness for (a-d) $N=1-4$ layers, resp.  Calculated bands for three
configurations are shown: Bernal stackings ABAB and ABAC (blue and light blue, resp.)  and Rhombohedral stackings (red).Adapted from
Ref.\ \cite{ohta2007}.  }
\def\figFLG{\begin{figure}\center{\includegraphics[width=5in]{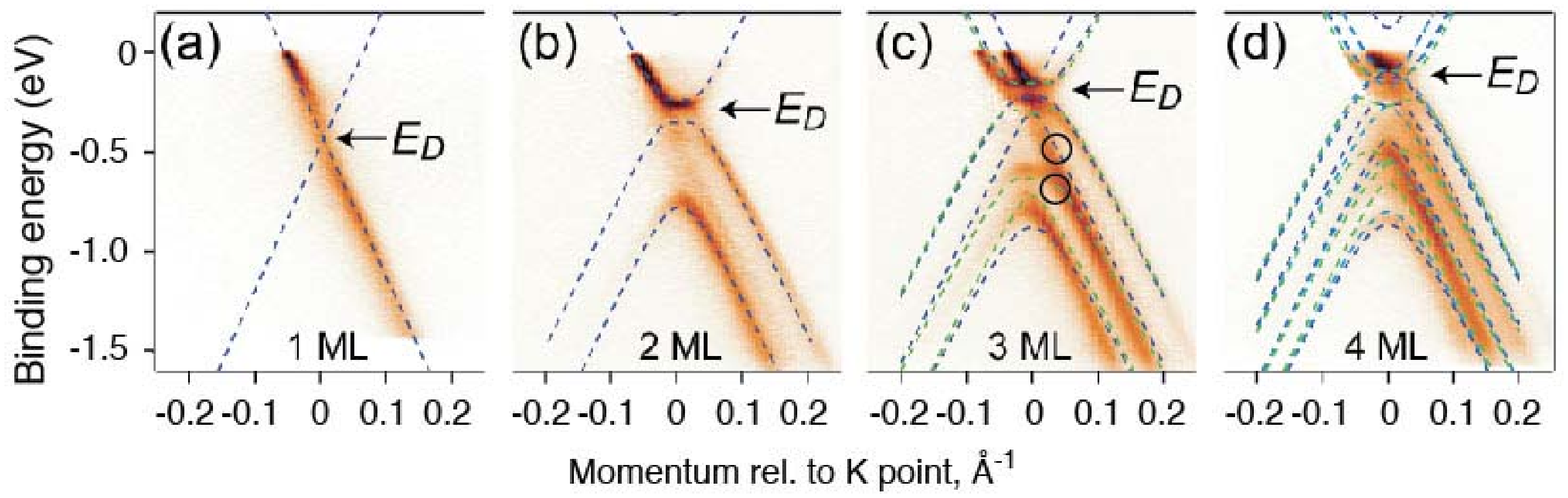}}\caption{\label{f:figFLG}\figFLGText}\end{figure}}

\title[Symmetry Breaking in FLG Films]{Symmetry Breaking in Few Layer Graphene Films}

\author{Aaron Bostwick$^{1}$, Taisuke Ohta$^{1,2}$, Jessica L. McChesney$^{1,3}$, Konstantin V. Emtsev$^{4}, $Thomas Seyller$^{4}$,
Karsten Horn$^{2}$, Eli Rotenberg$^{1}$}


\address{$^{1}$Advanced Light Source, E. O. Lawrence Berkeley National Laboratory, Berkeley CA 94720 USA}
\address{$^{2}$Department of Molecular Physics, Fritz-Haber-Institut der Max-Planck-Gesellschaft, Faradayweg 4-6, 14195 Berlin, Germany}
\address{$^{3}$Department of Physics, Montana State University, PO BOX 173840 Bozeman, MT 59717-3840 USA}
\address{$^{4}$\thomasAddr}
\ead{erotenberg@lbl.gov}
\begin{abstract}
Recently, it was demonstrated that the quasiparticle dynamics, the layer-dependent charge and potential, and the $c$-axis screening
coefficient could be extracted from measurements of the spectral function of few layer graphene films grown epitaxially on SiC using
angle-resolved photoemission spectroscopy (ARPES).  In this article we review these findings, and present detailed methodology for
extracting such parameters from ARPES. We also present detailed arguments against the possibility of an energy gap at the Dirac
crossing \ED.
\end{abstract}

\section{Introduction}\label{s:intro}

\subsection{Isolation of graphene} \label{s:graphene}

Recently, much attention has been given to the electronic and other properties of graphene.  Following the isolation and dramatic
transport measurements of individual graphene flakes by exfoliation\cite{novoselov2004,novoselov2005,zhang2005}, there has been an
explosion of theoretical and experimental interest in graphene.  Among the interesting properties found are the massless,
relativistic nature of graphene's carriers, and the impact of Berry's topological phase factor on the transport properties of single
and bilayer graphene.  Especially interesting from a technological point of view is the extremely long lifetime of carriers, due to
weak backscattering arising from their chiral nature \cite{ando1998, ando1998b}.  This chiral nature derives from special symmetry
properties of the graphene lattice.

Exploitation of these effects for electronic devices requires the precise and scalable control of graphene nanostructures, which
cannot as yet be achieved with exfoliated flakes.  Therefore, much attention has been given to the epitaxial growth of graphene on
various substrates.  Forbeaux \etal\ were the first to demonstrate that high-quality epitaxy of single and few-layer graphene (FLG)
could be achieved on the silicon-rich SiC(0001) surface\cite{forbeaux1998}.  Transport measurements and demonstration of the
feasibility of patterned graphene devices were demonstrated by Berger \etal\cite{berger2004,berger2006}.

\figAtoms

Fig.\ \ref{f:figAtoms} shows the atomic arrangement in monolayer and bilayer of graphene.  The unit cell consists of two equivalent
C atoms, labelled A and B with bond length $1.42\ \mathrm{\AA}$.  Jones proved that for a closed-packed hexagonal lattice, the
energy gap along the zone boundary disappears where bands from adjacent unit cells cross \cite{jones1934}.  This is illustrated in
Fig.\ \ref{f:figBand3d}, which shows the tight-binding (TB) band structure \Ek\ of graphene, evaluated to third nearest neighbor
using the parameters of Reich \cite{reich2002}.  (Here we restrict consideration to the $\pi$ and the \pistar\ states, which are
derived from the \pz\ orbitals of the carbon atoms\cite{saito1998}).  Quantitative fits of the TB model to experimentally determined
bands were presented by Bostwick \etal\ (Ref.\ \cite{bostwick2007b}).  These states meet at the so-called Dirac crossing point at
energy \ED\ in agreement with Jones' theorem.  For neutral (undoped) graphene, the Fermi energy (the energy of the least-bounds
states) \EF$=$\ED.

Many of the interesting properties of graphene revolve around the fact that the band crossing at \ED\ is strictly gapless.  This
means that at zero doping and zero temperature, graphene is a gapless semiconductor or a zero-overlap semimetal.  Upon doping the
graphene by either deposition of foreign atoms \cite{ohta2006, bostwick2007}, molecules \cite{wehling2007} or in a gated geometry
\cite{novoselov2004,novoselov2005,zhang2005}, the carrier density can be easily manipulated.  With this control, we can
systematically
study the many-body interactions in graphene as a function of doping using angle-resolved photoemission spectroscopy. 

\figBand3d

\subsection{Angle-resolved Photoemission Spectroscopy} \label{s:arpes}

The Fermi surface is defined as a constant energy surface \Ek$|_{E=E_{\mathrm{F}}}$, and determines all the transport properties of
conducting materials.  While transport measurements on doped graphene can determine the relevant properties such as group velocity
and lifetime of carriers on the Fermi surface, angle-resolved photoemission spectroscopy (ARPES) is a useful complementary tool.  It
can determine the electronic band structure, so it is capable of measuring not only the group velocity and Fermi surface, but also
the constant-energy surfaces for all occupied states and the full occupied bandstructure \Ek.  Furthermore, the technique also
accesses important information about many-body effects\cite{damascelli2003}.  When there is sufficient energy and momentum
resolution, the experimentally determined spectral width of the Fermi contours can be taken to be the inverse of the mean free path,
and the measurement of \Ek\ is taken as a measure of the many-body spectral function \akw.

This spectral function is in turn related to an electronic self-energy \skw\ as follows (see Ref.\ \cite{kaminski2005} and therein):

\begin{equation}
    \label{e:akw}
    \mathakw=\frac{\left|\mathimskw\right|}{\left(\omega-\omega_{\mathrm{b}}(\mathrm{\mathbf{k}})-\mathreskw\right)^{2}+(\mathimskw)^{2}},
\end{equation}
where $\omega$ is the measured binding energy and \wb\ is another energy defined below (where $\hbar=1$).  We make the approximation
that \skw\ is \textbf{k}-independent.  In this form, we see that \akw, when evaluated at constant energy $\omega$, is a Lorentzian function whose
width is given by \ims\ representing the inverse lifetime (proportional to the inverse mean free path).  

Eq.\ \ref{e:akw} is valid when the scattering rate of the charge carriers (expressed in energy units) is not too large compared to
their energy.  In this situation, we refer to the charge carriers as quasiparticles (QPs).  In our measurements, the QPs are holes
which have been injected as part of the photoemission process.  Their binding energy $\omega$ (here $\hbar=1$) is taken as a
negative number, and we speak of increasing energy as an increase in this negative value. 

One can draw an analogy between QPs propagating in a scattering medium and light traveling in a lossy optical medium.  Such a
medium is characterized by a complex dielectric function, and the effects on the light propagation are not only through its
absorption but also a dispersion.  To satisfy causality, the real and imaginary parts of the dielectric function are related by a
Hilbert transformation.  Similarly, the propagation of QPs in a scattering medium leads not only to inelastic scattering
(whose lifetime is encoded in \ims) but also renormalization of the carrier's energy, encoded in the real part of \skw.  These real
and imaginary parts of \skw\ are also related by a Hilbert transform, and the function \res\ is defined as the difference between
the measured carrier energy $\omega$ and the ``bare'' band energy \wb\ (that is, in the absence of scattering interactions), as
indicated in Eq.\ (1).  Following this formalism, ARPES can determine the energy-dependent lifetime due to scattering from other
excitations in the system.

For a valid spectral function analysis, the ARPES spectra must be acquired with sufficient resolution and the samples
must be of high quality so that defect scattering is negligible.  They must also be well-characterized in thickness to ensure that
the pure graphene signal is accessed.  

The first ARPES measurements on FLG on SiC were  from thick films\cite{strocov2001,kihlgren2002,soe2004} aimed toward studying the
properties of graphite.  Later, Rollings \etal\ \cite{rollings2006} measured the Fermi surface and other constant energy surfaces
around \ED\ for a film with thicknesses around 2-3 layers, determined by core level shifts of C $1s$ electrons.  Systematic core
level and valence band offset studies were carried out around the same time by Seyller \etal\cite{seyller2006}. Because of the
contribution of carbon from the SiC substrate to the core level signal, such measurements give a rough measure of the film thickness
but cannot give a precise thickness measurement.  

As shown below, the ARPES measurements themselves can give not only a precise thickness determination, but also determine other
crucial quantities.  The initial formation of the graphene valence band from the silicon-rich SiC surface through to the first
monolayer graphene was by Emtsev \etal\cite{emtsev2006}.  Valence band measurements to discriminate film thicknesses greater than
1 monolayer were first shown by Ohta \etal\ for bilayer\cite{ohta2006} and later systematically for monolayer-quadlayer graphene
films\cite{ohta2007}.

These studies also demonstrated the crucial role of substrate preparation for good quality valence band measurements. The first
 detailed spectral function by ARPES from graphene were published by Bostwick \etal\cite{bostwick2007} and could show a rich spectrum
dominated by electron-electron, electron-phonon, and electron-plasmon scattering.

\section{Experimental}

Here we briefly review the growth method of graphene on SiC in our work \cite{ohta2006,bostwick2007,ohta2007}. Films were grown on $n$-type ($\sim10^{18}$ cm$^{-3}$ N atoms) SiC(0001) wafers which were precleaned by annealling in 1 bar of
Hydrogen gas at 1550$^{\circ}$C for around 30 minutes.  The role of this cleaning step is essential, as by etching it removes the
polishing scratches while maintaining bulk SiC stoichiometry.  As-cleaned substrates were found to be atomically flat with wide
terraces (Ohta \etal, unpublished).  Formation of graphene layers by heating to around 1200$^{\circ}$ in ultrahigh vacuum was
monitored with low energy electron diffraction (LEED) following Forbeaux\cite{forbeaux1998} and ARPES as described below.  The base
pressure of our system was 1-$2\times 10^{-11}$ T, and graphene growth was always performed at pressures better than $1\times
10^{-10}$ T. All measurements were obtained at phonon energy $h\nu=$94 eV unless otherwise noted.

\section{Symmetry breaking considerations for few layer graphene}\label{s:symm}
\subsection{Monolayer Graphene: a gap at \ED\ due to symmetry breaking?}\label{s:ml}
As Forbeaux \etal\ showed, FLG formation is accompanied by a \sixr3\ reconstruction at the graphite-SiC interface, which was
initially attributed to the interference between the graphene and SiC lattice constants \cite{forbeaux1998}.  We now know from
photoemission\cite{emtsev2006}, theoretical calculations\cite{varchon2007,mattausch2007}, and scanning tunnelling microscopy (STM)
\cite{mallet2007} that the \sixr3\ represents a non-interacting ``\zeroth'' graphene layer whose electronic structure resembles
graphene only insofar as it has an intact $\sigma$-like bands (derived from $sp^{2}$-hybridized in-plane bonds) but lacking the
$\pi$ bands characteristic of the out-of-plane \pz\ states of graphene.  The presence of such a \zeroth\ layer is important because
it acts like a dead layer, saturating or interacting with the underlying SiC bonds while forming a template for a subsequent first
graphene overlayer.  From symmetry considerations, it is known that the $\pi$ bands from the latter and the $\sigma$ bands of the
former cannot interact.  Therefore, the first graphene layer's chemical interaction with the substrate is very weak, and therefore
we expect the $\pi$ bands of graphene on SiC are to a very good approximation the same as those of freestanding doped graphene.  In
the following, we do not count this dead \zeroth\ layer among the active graphene layers in our FLG film.

It is well-known that the Hamiltonian of one layer of graphene near the K point of the Brillouin zone can be approximated
\cite{ando1998, ando1998b,divincenzo1984,mccann2006} by

\def\tr{\mathrm{T}} \def\ai{\alpha_{i}=\left( \begin{array}{cc} E_{i} & v\pi^{\dag} \\ v\pi &
E_{i} \end{array} \right)} \def\bi{\beta_{s}=\gamma_{1}\left( \begin{array}{cc} 0 & s \\ 1-s & 0 \end{array} \right)}
\begin{equation}\label{e:H1}
  {\cal{H}} = \left( 
     \begin{array}{cc} E_{1}+\Delta/2 & v(k_{x}-ik_{y}) \\ v(k_{x}+ik_{y}) & E_{1}-\Delta/2\\    \end{array}
    \right) \equiv
     \left(
     \begin{array}{cc} E_{1} & v\pi^{\dag} \\ v\pi & E_{1}\\    \end{array} 
     \right) \equiv
    \alpha_{1}
\end{equation}
where the wavefunctions $\Psi=(\psi_{\mathrm{A}}, \psi_{\mathrm{B}})$ are written in terms of \pz\ orbitals centered on the A and B
atoms in the graphene basis set.  The parameter $\Delta$ represents a possible asymmetry between the A and B sites.  For ordinary
graphene, $\Delta=0$ since the atoms are indistinguishable.  The off-diagonal terms represent the hopping between A and B
sublattices, and $v$ is the band velocity around \ED.

\figExptML

The Hamiltonian in Eq.\ \ref{e:H1} leads to a conical bandstructure \Ek$=vk$ when $\Delta=0$.  Here $\mathbf{k}$ is the momentum
relative to one of the K points at the corner of the graphene Brillouin zone (see Fig.\ \ref{f:figBand3d}).  Experimental Fermi
surfaces and underlying bandstructures for clean and alkali-dosed graphene are shown in Fig.\ \ref{f:figExptML}(a-b), adapted from
Ref.\ \cite{bostwick2007}.  We can immediately recognize the expected nearly linear dispersions as well as the Dirac crossing points
(middle panels) in the bands at the Dirac energy \ED. We also see that there is a non-trivial change in intensity when traversing
around the Fermi contour.  This will be discussed in detail below, but for now we regard it as a photoemission cross section effect.
Because of this effect, when we sample the bandstructure in the $y$-direction (relative to Fig.\ \ref{f:figExptML}), we see only one
of the two expected bands; the other is extinguished (right panels).

We also observe that even the clean, as-grown graphene films have a Fermi level \EF\ significantly above (by around 0.45 eV) the
Dirac energy \ED. This in-built doping was first reported by Rollings \etal\cite{rollings2006} and can be attributed to the greater
electron affinity of graphene compared to the substrate.  Our experiments have shown that this intrinsic $n$-doping is independent
of whether the substrate dopants have been frozen out (at $T\sim 40$K).  Since its discovery by ARPES, this intrinsic $n$-doping
has also been predicted theoretically \cite{varchon2007,mattausch2007}.

An important feature of the one monolayer data is the appearance of kinks in the energy band structure below \EF\cite{bostwick2007}.
These kinks occur at two energy scales.  First we see a slight kink at $\sim 200$ meV below \EF. This kink is hardly visible on the
large energy scale plotted in Fig.\ \ref{f:figExptML}, but it is accompanied by pronounced sharpening between 200 meV and \EF\ that
is readily observed.  This kink is similar to ones which have been observed at similar energy scale in
graphite\cite{zhou06,sugawara07} and bilayer graphene \cite{ohta2006} that have been attributed to electron-phonon scattering.  We
can understand the presence of the kink within the spectral function formalism in Eq.\ \ref{e:akw}, noting that there is an
observable increase in linewidth of the band at binding energies greater than 200 meV, signifying a decrease in the lifetime of the
states as electrons absorb or emit phonons.  We will discuss this formalism further below but for now it is sufficient to identify
this feature with phonons for two reasons: first because of the energy scale, which corresponds to the in-plane LO and TO phonons,
and second, because the kink feature's energy scale remains constant with doping, as expected for the small doping levels considered
here.

There is a second anomaly in the dispersion around the Dirac crossing point in Fig.\ \ref{f:figExptML}.  In the middle panels, where
both bands have equal intensity, the region of the crossing of the bands seems spread out in energy.  In the right panels, where one
band is extinguished, it appears that this spread is associated with a second kink feature which is at the Dirac crossing point of
the bands.  Unlike the phonon kink, this anomaly moves to higher binding energy with doping, and must therefore be somehow
associated with the Dirac energy \ED. Similar to the phonon kink, it is stronger at higher doping, and it is associated with a change
in linewidth--the bands are locally broadened around \ED.

\figSat

What causes this second feature?  In Bostwick \etal, it was proposed to be a kink due to electron-plasmon
interaction\cite{bostwick2007} but it has been recently proposed that the observed spreading of the bands around \ED\ is associated
with substrate-dependent energy gap at \ED \cite{lanzara2007,zhou2007}.  Such a gap would be interesting because it suggests an
electronic or chemical control of the electronic character (2D semimetal \textit{vs}.\ semiconductor) and is proposed on the basis
of possible symmetry breaking.  First, we discuss this idea and then present the evidence against it followed by evidence in favor
of the electron-plasmon scattering model.

Within the simple Hamiltonian (Eq.\ \ref{e:H1}) a gap of magnitude $\Delta$ appears at the Dirac crossing energy \ED\ when the
parameter $\Delta\neq 0$.  A physical interpretation of this gap is the symmetry breaking of the A and B atoms.  This occurs for,
e.g. replacement of C atoms with B and N in hexagonal boron nitride.  It also occurs in a scenario where the bonding of A and B
atoms to the \zeroth\ layer is asymmetric as recently proposed\cite{lanzara2007,zhou2007}.

We present arguments against this scenario in graphene on SiC as follows.

(1) The interaction between the 1st and 0th layer is very weak.  This was established by ARPES\cite{emtsev2006},
theory\cite{varchon2007,mattausch2007}, and STM\cite{mallet2007}.  A possible argument against the weak interlayer attraction is the
appearance in monolayer graphene films of replica $\pi$ bands with \sixr3\ symmetry, ascribed by us as due to diffraction of the
outgoing photoelectrons\cite{bostwick2007}, similar to other nearly incommensurate systems\cite{rotenberg2003}.  These satellite
bands lead to replicas of the constant energy contours, illustrated for the Dirac crossing energy in Fig.\ \ref{f:figSat}.  With a
linear grey scale in (a) the replica bands are hardly noticable but with a highly non-linear grey scale (b), they can be
emphasized.  

It is tempting to ascribe the replica bands to a possible \sixr3\ superlattice potential felt by the first graphene layer.  If this
were true, additional energy minigaps would appear where the replica and main bands cross\cite{crain2002};however, no such gaps have
been observed\cite{ohta2006}.  Furthermore, the replica bands, very weak at low temperature (about a factor 40
reduced intensity compared to the primary band) do not appear at room temperature\cite{emtsev2006} and at 100K are dramatically
broadened [J. L. McChesney, unpublished].  This observation violates the hypothesis that the first graphene layer has \sixr3\
symmetry potential which would demand the linewidths of the replica and main bands to be identical by symmetry.  Instead, we can
easily understand the broadening of the replica bands as due to a Debye-Waller factor, confirming the origin of these replica bands
as due to final-state diffraction.

(2) The doping dependence shows a clear increase in the spread of the states at the Dirac crossing.  If this spread were due to a
gap from coupling to the substrate, the coupling strength should be independent of the doping density (or become smaller due to
enhanced screening).

(3) We observed that the bands above and below \ED\ are misaligned\cite{bostwick2007}, so that the projections of the $\pi$ states
below \ED\ do not pass through the $\pi^{*}$ states above \ED.  This is illustrated by the dashed lines in
Fig.\ \ref{f:figNoGap}(a), which reproduces the clean graphene bandstructure.  This misalignment does not occur in the energy gap
scenario, but comes naturally when many-body interactions are present.

(4) The density of states (DOS) does not show a gap at \ED. This is illustrated in Fig.\ \ref{f:figNoGap}(b) for the momentum-integrated
DOS. In a gap scenario one expects a decreased DOS but we see a peak (expected for crossed bands).

\figNoGap

(5) The energy distribution curve (EDC) at the Dirac crossing shows only a single peak, not a split peak as expected in a gap
scenario (see Fig.\ \ref{f:figNoGap}(c)).  

(6) The intensity distribution along the Fermi surface provides a stringent test for A-B atom symmetry breaking.  It is observed
that one side of the Fermi contours is very weak or absent.  In the strictly symmetric case $\Delta=0$, the intensity on one side of
the Fermi contour is strictly zero.  Rather than a simple vanishing photoemission matrix element, the cancellation results from the
interference between emission from A and B sites, as shown by Shirley\cite{shirley1995}.  This cancellation, like the Dirac nature
of the quasiparticles, and the lack of backscattering, follow from the strict A-B atom symmetry.  If we break the A-B atom symmetry,
we not only open a gap at \ED\ (thereby destroying the massless character), but also destroy the phase cancellation affecting the
Fermi surface intensity.

These effects are illustrated in Fig.\ \ref{f:figNoInt}.  In (a) we show as polar maps the measured angular distribution of the band
intensity taken about the K point for monolayer and bilayer graphene (closed and open circles, resp.).  These data were obtained by
fitting the momentum distribution curves taken along radial cuts for an energy window $\sim 75$ meV below \EF. The monolayer and
bilayer Fermi surfaces are practically identical, but as indicated in the figure, the bilayer signal is not completely extinguished
in any direction.  In contrast, for the monolayer, the intensity is completely extinguished in one direction, apart from a
very weak minority contribution from bilayer regions.  This residual bilayer signal is easy to subtract since it is well
separated from the monolayer bands below \ED\cite{ohta2006}.  After subtraction, we determined that the monolayer band
intensity is zero within a very low noise floor (about 0.15\% as indicated by the central yellow circle).

\figNoInt

Shirley derived a simple formula for the symmetric case $\Delta=0$ for monolayer graphene; we extended this model to the case of
finite $\Delta$ and show in Fig.\ \ref{f:figNoInt}(a) the expected angular distributions for a $\Delta=0.0, 0.1, 0.2$ eV (leading to
energy gaps at \ED\ of the same values).  This plot shows that we are fairly sensitive to the possible symmetry breaking (and this
sensitivity can be enhanced simply by acquiring the bands with better statistics).  Fig.\ \ref{f:figNoInt}(b) shows a plot of the
intensity reduction as a function of $\Delta$, which can be compared to our noise floor ($<.015\%$).  From this comparison, we can
conservatively estimate the maximum gap at \ED\ to be under 60 meV. Since the apparent kink at \ED\ (with a resulting spreading of
the states there) is much wider in energy than this, we can rule out the symmetry breaking as being the dominant factor to explain
the anomalous dispersion at \ED.

As an aside, the reason the bilayer is not completely extinguished is that A-B atom symmetry breaking is indeed violated for the
bilayer.  That is because only one atom (A, say) in the outer layer is directly over an atom in the inner graphene layer (see Fig.\
\ref{f:figAtoms}).  This symmetry breaking also explains the well-known symmetry breaking in STM images of bilayer and thicker
films\cite{mallet2007,shirley1995}.  (A complete model of the bilayer angular intensity profile is outside the scope of this
paper and will be presented elsewhere.)

(7) It is worth pointing out the very high momentum resolution and accuracy of positioning of the sample that is required to obtain
spectra precisely at \ED.  In Fig.\ \ref{f:figNoGap}(a), we see that the entire span of the Fermi bands is only about 0.1 $\mathrm{\AA}^{-1}$.
Only a small misalignment on the order of 0.05$^{\circ}$ could result in an apparent gap in the bands.

\subsection{Many-body explanation for anomalies at \ED}\label{s:mb}
\subsubsection{The case for self-consistency.}\label{s:ep}

\figSpecComp

Having ruled out the gap scenario, we can now consider many-body interactions to explain the kinked dispersions around \ED. The
first issue is whether a self-consistent model is possible even in principle.  We will first establish that the kinks and the
linewidth variations are consistent with each other.  As discussed above and in the literature\cite{kaminski2005,kordyuk2005}, we
analyze the spectral function data in terms of real and imaginary parts of the self-energy function \skw.  Fig.\
\ref{f:figSpecComp}(b) shows an experimentally acquired spectral function \akw\ for relatively highly doped graphene (\n56).  The
dressed band position $\omega=$\wb + \reskw\ is determined by fitting momentum distribution curves (MDCs, that is, individual
constant-energy slices) to Lorentzian functions.  The positions are plotted in Fig.\ \ref{f:figSpecComp}(a) (black line) and the
Lorentzian width as a function of $\omega$ is plotted in Fig.\ \ref{f:figSpecHilbert}(a).

\figSpecHilbert

In order to converge to a self-consistent interpretation, it is necessary to iteratively apply the following procedure.  We take a
second order polynomial as a trial bare band \wb.  Given this \wb, we can easily scale the MDC widths (units of $\mathrm{\AA}$) into
the function \imskw\ (units of eV), shown in \ref{f:figSpecHilbert}(b).  This function is smoothed and then Hilbert transformed into
a trial \reskw\ function.  We can also extract the \reskw\ function by subtracting the trial bare band from the fitted band
position.  These two \reskw\ functions (Fig.\ \ref{f:figSpecComp}(c)) are compared and the trial bare band adjusted until the model
\reskw\ and \imskw\ are in good agreement with the experimentally extracted curves as plotted in Fig.\ \ref{f:figSpecComp}(b-c).

As a final check of self-consistency, we can generate a trial spectral function \akw\ from only the fitted MDC widths and the
mathematically transformed \reskw, shown in Fig.\ \ref{f:figSpecComp}(b).  It is in excellent agreement with the experimental
function in Fig.\ \ref{f:figSpecComp}(a).  Having demonstrated this self-consistency, we can say with high degree of confidence that
the kink anomalies must be attributed to many-body interactions, and not any details of the single-particle bandstructure.  That is,
we can safely rule out not only the superlattice gap scenario outlined above, but also strain, defects and other initial-state
effects.

\figSpecTheory
\subsubsection{Contributions to scattering lifetime.}
\figProcess We now explain the physical origin of the measured \imskw\ function in more quantitative fashion.  For convenience we
work with the fitted MDC widths, which are plotted in Fig.\ \ref{f:figSpecTheory}(a).  The features to explain are, starting from
zero energy, the monotonic increase in lifetime down to about $-0.2$ eV; the hump at around \ED$=-1.0$ eV, and the remaining background
rise.  These were attributed \cite{bostwick2007} to electron-phonon (\eph) coupling, electron-plasmon (\epl) coupling, and
electron-hole (\eh) pair generation; computations of these contributions to the scattering rate are shown in Fig.
\ref{f:figSpecTheory}(b).  Schematic diagrams of these processes are shown in Fig.\ \ref{f:figProcess}.

We can meaningfully consider only those excitations whose energy scale is greater than our energy resolution ($\sim 25$ meV).
Considering the energy scale of the observed kink anomalies ($\geq 200$ meV) we can rule out any significant interactions between 25
to 200 meV, such as scattering from low-energy acoustical vibrational modes.

First, we qualitatively discuss the coupling to phonons at the 200 meV energy scale (a quantitative analysis has been presented by
Bostwick \etal\cite{bostwick2007b} for graphene, and for other surfaces in Refs.\ \cite{valla1999, hengsberger1999, rotenberg2000}).  Since this energy is much larger than our temperature
($k_{\mathrm{B}}T\sim 2$ meV, we can rule out phonon absorption and consider only decay of quasiparticles (QPs) by phonon emission
(Fig.\ \ref{f:figProcess}(a)).  Such QP decays are forbidden for states at \EF, but become available as the quasiparticle energy
increases.  Typically once the quasiparticle energy is greater than the phonon energy scale, the lifetime due to scattering is
independent of QP energy.  This change in QP lifetime is reflected in a monotonic increase in the imaginary part of the self energy
\imskw.  Because the real and imaginary parts of \skw\ are related by Hilbert transform, one expects to see a non-trivial change in
the dispersion on the phonon energy scale, which is observed as a kink.  Physically, we interpret the change of band slope between
the kink and \EF\ as a renormalization of the mass as the QPs become ``dressed'' with a virtual cloud of phonons.  But we know that
the QPs in graphene are effectively massless, so we speak in terms of a velocity renormalization (or equivalently a renormalization
of the relativistic mass-equivalent energy).

The 200 meV kink is stronger for the K-covered graphene compared to the as-grown surface (see Fig.\ \ref{f:figExptML}(a,b)) due to a
phase-space argument.  The density of electronic states in \kspace\ is a constant, so that as the sample is doped, the bands span
more electronic states near \EF; as these become available final states for phonon scattering, the decay probability increases.
Left unexplained is the overall magnitude of the \eph\ scattering rate at all dopings, which is about 6 times stronger than what is
predicted by the normal deformation potential calculations \cite{bostwick2007b,calandra2005,calandra2006}.

The quantitative analysis of the phonon kink \cite{bostwick2007b,bostwick2007}, which followed the standard formalism
\cite{grimvall1981}, is quite useful but does not perfectly describe the kink strength (it underestimates it slightly) and
furthermore does not take into account the actual band structure of graphene: the actual phonon scattering rate should diminish
near \ED\ from the same phase-space argument just cited.  A first-principles calculation of the phonon scattering rate should
account for both of these effects.

In the case of the second kink near \ED, the QP decay is through emission of plasmons (Fig.\
\ref{f:figProcess}(b)), which is favored over phonon scattering because of the kinematic
constraints\cite{bostwick2007,vafek2006,hwang2006b}.  Whereas optical phonons are more or less delocalized in \kspace\ with fixed
energy scale, the plasmon spectrum in graphene has a fast energy dispersion in a narrow range of \k.  This follows from the
dispersion relation for two-dimensional plasmons \cite{stern1967} in the long wavelength limit:

\begin{equation}
    \label{e:pl}
    \omega_{\mathrm{pl}}(q)=\sqrt{4\pi n e^{2}q/m(1+\epsilon)}
\end{equation}
where $q$ is the plasmon momentum, $m$ is the carrier mass, and $\epsilon\sim 6$ is the dielectric
constant\cite{bostwick2007b}.  Although plasmons in principle exist in the domain $0 < q < \infty$, in practice they
propagate freely up to a critical momentum $q < q_{\mathrm{c}}$ due to Landau damping (plasmon decay into electron-hole
pairs)\cite{kliewer1973}.

For graphene, the rest mass $m$ is zero near \ED\, but the relativistic mass equivalent to the kinetic energy, \mr=$E/v^{2}$ (where
$v$ is the Fermi velocity), is on the order\cite{novoselov2005,zhang2005} of 0.1\me\ and can be used to set the plasmon energy scale \wpl. 
This means that more or less vertical interband decays by plasmon scattering are now the dominant factor determining the lifetime
near \ED.  

Two other contributions to the scattering lifetime must be considered. First is defect or impurity scattering.  Normally the defect 
scattering is taken to be a constant background to the imaginary self energy \imskw, which is a deconvolution of the residual
momentum spread of the bands at \EF and the instrumental function. In our case, the residual momentum spread is only about $0.005\ 
\textrm{\AA}^{-1}$, which is comparable to our instrumental resolution, so we can safely discard the defect scattering rate as
negligible.

The remaining contribution to the scattering rate is the decay by \eh\ pair generation, which is the standard decay process in Fermi
liquid theory.  In this process (Fig.\ \ref{f:figProcess}(c)), the decay of the quasiparticle is accompanied by an excitation
of an electron above \EF, creating a new hole below \EF. For two-dimensional metals with circular Fermi surface, Hodges \etal\ proved the famous rule that \eh-pair
scattering rate goes as $\omega^{2}$ln$\omega$\cite{hodges1971}.  This was determined by a phase space integration of all possible
kinematically allowed \eh-scattering processes.  For a 2D free electron gas this could be carried out analytically, but for
graphene, we evaluated the appropriate integral numerically.  This was done so that we could use the experimentally determined
dispersion (although we assumed cylindrical symmetry and zero temperature to simplify the integration).

The most interesting finding is that for $n$-doped graphene, the \eh-scattering rate rises from \EF\ down towards \ED\ as it would
be expected for any metal.  Around \ED, however, the scattering rate must necessarily drop, because in the vicinity of \ED, the
decays are mostly vertical transitions.  Such a decay by \eh-pair generation is forbidden since we cannot find a momentum-conserving excitation 
near \EF\ to satisfy the kinematic constraints.  Only at energy scale around twice the Dirac energy do such excitations become
available, and we see an associated rise in the scattering rate at high energy scales.

Considering the simplicity of the model, the total scattering rate function (Fig.\ \ref{f:figSpecTheory}(b)) does a remarkably good
job to model the data.  Theoretical modelling of the \epl\ and \eh\ scattering rates has also been performed by Hwang
\etal\cite{hwang2006b}.  Fig.\ \ref{f:figSpecTheory}(c-e)) shows a comparison of our measured MDC widths to their model for three
different dopings.  Although they overestimate the relative contribution of the \epl\ to \eh\ processes, their calculation is in
excellent qualitative agreement with the observed MDC widths.  The main discrepancy is the failure of the model to account for the
scattering rate from phonons, which was not included in their calculation.

The many-body effects we measure  are present all the way down to zero doping and therefore may play a role in the
transport of gated graphene devices.  These are much more dilute carrier gases than we achieved by alkali metal doping.  As the
doping level decreases, the phonon and plasmon processes will overlap in in energy and therefore will not be separable.  This is
already seen in the lowest doping we probed (Fig.\ \ref{f:figSpecTheory}(c)).  The plasmon and \eh\ pair scattering rates are
reasonably separable at all dopings, but there is an energy overlap region just above \ED\ where neither alone is a good description
of the total electron-electron interaction.  These observations imply that a proper description of the scattering rate will take
into account much more complicated processes than in our simple treatment.  In the language of Feynman diagrams, it means
higher-order diagrams than are typically considered will be necessary to model the photoemission data.  In addition, when \EF\ is
reduced to be comparable to the temperature, thermal excitation effects will increase in importance.  This has already been
discussed in relation to plasmons \cite{vafek2006}.

\subsection{Out-of-plane symmetry breaking in multilayer graphene.}\label{s:multi}

Multilayer graphene films grown on Silicon carbide have an obvious built-in symmetry breaking, because of the inequivalence of the
two film interfaces (SiC and vacuum).  Further symmetry breaking can be induced by either external field, or by growth of additional
layers on top of the graphene films.  Understanding these symmetry effects is important in order to exploit them for technological
purposes.  Extension of the Hamiltonian in Eq.\ \ref{e:H1} to multiple layers gives a simple framework to achieve this.

Extension to two layers is achieved by adding an additional hopping term between the B atoms of the first layer and the A atoms of
the second layer \cite{mccann2006b, guinea2006, ohta2007}:

\begin{equation}\label{e:H2}
   \cal H =\left( 
    \begin{array}{cccc} 
	E_{1}      & v\pi^{\dag} & 0          & 0          \\
	v\pi       & E_{1}       & \gamma_{1} & 0          \\
	0          & \gamma_{1}  & E_{2}      & v\pi^{\dag} \\
	0          & 0           & v\pi       & E_{2}
    \end{array}
    \right) \equiv
    \left(
    \begin{array}{cc} \alpha_{1} & \beta_{0} \\ \beta_{0}^{T} & \alpha{2} \end{array}
    \right)
\end{equation}
Here $\alpha_{i}$ acts with respect to the (A, B) sublattices of the $i^{th}$ layer, and $\beta_{0}$ is a $2\times 2$ matrix

\begin{equation} \beta_{0}\equiv
    \cal H=\left(
    \begin{array}{cc} 0 & 0 \\ \gamma_{1} & 0 \end{array}
    \right)
\end{equation}
where \g1\ is the hopping parameter between layers.  The wave function now has four elements with basis set orbitals located at
A$_1$, B$_1$, A$_2$, B$_2$ atoms, where i is the layer number (1 or 2).

There are two further generalizations of Eqn.\ \ref{e:H2}.  First, by adding more layers, and second by altering the stacking 
sequence.  Adding a third layer, one couples the B atom of the second layer to the A atom of the third for conventional Bernal-type 
stacking ($ABAB\ldots$) characteristic of bulk graphite. Repeating this sequence we come to the Bernal Hamiltonian for $N$ layers, 
\begin{equation}\label{e:HN}
   \cal H=\left(
    \begin{array}{cccccc}
	\alpha_{1}      & \beta_{0}        &                 &                     &                  &                    \\
	\beta_{0}^{\tr} & \alpha_{2}       & \beta_{0}^{\tr} &                     &                  &                    \\
			& \beta_{0}        & \alpha_{3}      & \beta_{0}           &                  &                    \\
			&                  & \beta_{0}^{\tr} & \alpha_{4}          & \beta_{0}^{\tr}  &                    \\
			&                  &                 & \beta_{0}           & \ddots           &                    \\
			&                  &                 &                     &                  &\alpha_{\mathrm{N}} \\
	\end{array} \right)   \hspace{0.5in}  \textrm{(Bernal),}
\end{equation}

A useful generalization of Eq.\ \ref{e:HN} is
 \begin{equation}\label{e:Hgen}
     \cal H = \left( \begin{array}{cccccc}
     \alpha_{1}      & \beta_{0}        &                 &                     &            &                    \\
     \beta_{0}^{\tr} & \alpha_{2}       & \beta_{s}       &                     &            &                    \\
		     & \beta_{s}^{\tr}  & \alpha_{3}      & \beta_{0}           &            &                    \\
		     &                  & \beta_{0}^{\tr} & \alpha_{4}          & \beta_{s}  &                    \\
		     &                  &                 & \beta_{s}^{\tr}     & \ddots     &                    \\
		     &                  &                 &                     &            &\alpha_{\mathrm{N}} \\
     \end{array} \right) \hspace{0.5in} \textrm{(General),}
 \end{equation}
where
\begin{equation} \bi.
\end{equation}
Now, if $s=0$, then Eq.\ \ref{e:Hgen} is the Hamiltonian for Bernal stacking, while for $s=1$, Eq.\ \ref{e:Hgen} is the Hamiltonian 
for Rhombohedral stacking $(ABCABC\ldots)$.  We can further generalize this Hamiltonian to arbitrary stacking orders, by suitably 
choosing the different values of $s$ in each block of the matrix.

In the above Hamiltonians, we have assigned to each layer  its own on-site Coulomb energy \Ei.  This
allows for the possibility of a poorly screened electric field across the FLG film, which is reasonable in view of the predicted
long screening lengths in this direction.  It is straightforward to show that the Dirac crossing energy is \ED=Tr $\cal H$$/2N$ where
$N$ is the number of layers.

\figTheory

Fig.\ \ref{f:figTheory} shows the calculated bandstructures for one to four layer graphene.  The calculations were for either Bernal
(solid lines) or rhombohedral (dashed lines); the distinction is obviously meaningful only for films with $N\geq 3$ layers.  Far
from \ED, it turns out there is little distinction between rhombohedral and Bernal stacking.  This is to our advantage, because as
Fig.\ \ref{f:figTheory} shows, one can know the film thickness directly from band structure measurements by simply counting the
number of $\pi$ bands below \ED. Near \ED, the situation is quite different, since the two stacking types have quite different band
dispersions.  (Similar calculations have also been carried out with \emph{ab initio}
models\cite{min2007,wang2006b,varchon2007,mattausch2007,aoki2007}).

The detailed bandstructure around \ED\ shows a strong sensitivity to the Coulomb energy terms $E_{i}$ that enter the Hamiltonian
matrix (Eqs.\ \ref{e:H1}, \ref{e:Hgen})\cite{mccann2006}.  This can be seen by comparing the
upper and lower rows of Fig.\ \ref{f:figTheory} which were calculated for two cases.  In the first case, the energies $E_{i}$ are
all zero, and we find a gapless energy spectrum at \EF=\ED. For the lower row, we distributed a field change $U$=200 meV across the
total film in uniform increments, which simulates FLG in a bias or inhomogeneosly doped geometry.  This procedure opens gaps near
\ED; for the special bilayer case $N=2$, there is a complete gap at the Fermi level.  This gap was proposed to be the basis of a new
kind of electronc switch, whereby lateral transport through the bilayer could be modulated by applying a modest field perpendicular
to the film\cite{ohta2006,nilsson2006,castro2006}.

\figTwoLayer

Systematic studies of the thickness and doping dependence by ARPES have been presented by Ohta \etal\cite{ohta2006,ohta2007}. 
Fig.\ \ref{f:figTwoLayer} shows the bilayer graphene bandstructure in two different field geometries, achieved by doping a bilayer
graphene on SiC \cite{ohta2006}.  Simlar to monolayer graphene, the as-grown bilayer films have an intrinsic $n$-doping, which
allows us to probe the states both above and below \ED.  Because the doped carriers are concentrated in the interface layer, the
as-grown films have a field gradient across them and hence a gap at \ED.  Carefully balancing this charge imbalance allows us to
close the gap (Fig.\ \ref{f:figTwoLayer}(a)), while further doping of the surface layer allows us to create a net charge imbalance, 
thus reopening the gap (Fig.\ \ref{f:figTwoLayer}(b)). Evidence for a similar gap opening was also presented for the surface layers 
of graphite when doped with Na\cite{pivetta2005}.

Systematic thickness measurements at constant doping were presented by Ohta \etal\cite{ohta2007}.  For films of thickness $N$=1-4
layers, we found that the total charge density donated from the substrate was the same for all thicknesses.  Similar to bilayer
graphene, the charge was donated predominantly to the interface graphene layer.  This is in accord with the metallic nature of the
films, which can screen the interface layer from the rest of the film.  The measured bandstructures for $N=1-4$ layers are shown in
Fig.\ \ref{f:figFLG}.  

These spectra are very rich in information: we could determine not only the number of layers (by counting the number of
$\pi$ states below \ED\ straightforwardly) but also derive the stacking order.  One can see easily for $N$=3 that there are two
states (marked by circles) of equal intensity that can only be ascribed to equal populations of Bernal and Rhombohedral stacking. 
For quadlayer (and presumably thicker) careful analysis shows that Bernal stacking predominates. This can be taken as evidence for
the role of second near-neighbor interactions to stabilize the Bernal stacking type in graphite.

The electronic information that could be derived from the data are equally rich: in analogy with the bilayer analysis, we could
assign the different charge densities in each graphene layer, and determine the out-of-plane screening length.  In the formation of
the graphene films, about $1\times 10^{13}$ carriers per $cm^{2}$ are donated to the film, with in general about 85\% of the charge
donated to the first layer, and most of the rest in the second layer\cite{ohta2007}.

\figFLG

\section{Conclusions and outlook: Graphene, the simplest complex material.}\label{s:conc}

In the last few years, there was an explosion of interest in graphene since isolation of high-quality samples was achieved and since
its many novel properties were elucidated both experimentally and theoretically.  Seldom does a new material come along that has
such strong fundamental and practical interest.  From an experimental point of view, graphene is highly attractive since unlike
other low dimensional materials (such as high-mobility semiconductor two dimensional electron gases), graphene films are exposed to
vacuum and can be directly probed by surface-sensitive techniques such as LEED, STM, and ARPES. ARPES has a special role to play
because it is sensitive not only to the valence band energy structure but also its symmetry in $\mathbf{k}$-space.  Furthermore it
can give direct information on the many-body interactions, such as mass renormalization.  Through graphene's special sensitivity to
symmetry, we could even derive much structural information such as stacking errors and electronic information such as charge density
and screening length, which would be very hard to achieve with other probes.

In our opinion, graphene is unique in many ways.  It is the first system to our knowledge to show electron-plasmon coupling in the
ARPES signal, which suggests not only the exciting possibility of new coupling mechanisms, but also technological implications in
the interaction with photons.  Finally, it is a model system for correlation and many-body interactions which can supply stringent
tests for theory.

\label{refs}
\section*{References}

\begin{thebibliography}{10}

\bibitem{novoselov2004}
K.~S. Novoselov, A.~K. Geim, S.~V. Morosov, D.~Jiang, Y.~Zhang, S.~V. Dubonos,
  I.~V. Grigorieva, and A.~A Firsov.
\newblock Electric field effect in atomically thin carbon films.
\newblock {\em Science}, 306:666, 2004.

\bibitem{novoselov2005}
K.~S. Novoselov, E.~McCann, S.~V. Morosov, V.I. Fal'ko, M.~I. Katsnelson,
  U.~Zeitler, D.~Jiang, F.~Schedin, and A.~K. Geim.
\newblock Two-dimensional gas of massless dirac fermions in graphene.
\newblock {\em Nature}, 438:192--200, 2005.

\bibitem{zhang2005}
Y.~Zhang, Y.~W. Tan, H.~L. Stormer, and P~Kim.
\newblock Experimental observation of the quantum hall effect and berry's phase
  in graphene.
\newblock {\em Nature}, 438:201--204, 2005.

\bibitem{ando1998}
T~Ando and T~Nakanishi.
\newblock Impurity scattering in carbon nanotubes - absence of back scattering.
\newblock {\em J. Phys. Soc. Jpn.}, 67(5):1704--1713, 1998.

\bibitem{ando1998b}
T~Ando, T~Nakanishi, and R~Saito.
\newblock Berry's phase and absence of back scattering in carbon nanotubes.
\newblock {\em J. Phys. Soc. Jpn.}, 67(8):2857--2862, 1998.

\bibitem{forbeaux1998}
I.~Forbeaux, J.~M. Themlin, and J.~M. Debever.
\newblock Heteroepitaxial graphite on 6h-sic(0001): Interface formation through
  conduction-band electronic structure.
\newblock {\em Phys. Rev. B}, 58(24):16396--406, 1998.

\bibitem{berger2004}
C.~Berger, Z.~Song, T.~Li, X.~Li, A.Y. Ogbazghi, R.~Feng, Z.~Dai, A.~N.
  Marchenkov, E.~H. Conrad, P.~N. First, and W.~A. deHeer.
\newblock Ultrathin epitaxial graphite: 2d electron gas properties and a route
  toward graphene-based nanoelectronics.
\newblock {\em J. Phys. Chem. B}, 108(52):19912--19916, 2004.

\bibitem{berger2006}
C~Berger, Zhimin Song, X~Li, X~Wu, N~Brown, C~Naud, D~Mayou, T~Li, J~Hass,
  AN~Marchenkov, EH~Conrad, P.~N. First, and W.~A. de~Heer.
\newblock Electronic confinement and coherence in patterned epitaxial graphene.
\newblock {\em Science}, 312:1191--6, 2006.

\bibitem{jones1934}
H.~Jones.
\newblock Applications of the bloch theory to the study of alloys and of the
  properties of bismuth.
\newblock {\em Proceedings of the Royal Society of London. Series A,
  Mathematical and Physical Sciences (1934-1990)}, 147(861):396--417, 1934.

\bibitem{reich2002}
S.~Reich, J.~Maultzsch, C.~Thomsen, and P.~Ordejón.
\newblock Tight-binding description of graphene.
\newblock {\em Phys. Rev. B}, 66(3):035412, 2002.

\bibitem{saito1998}
R~Saito, G~Dresselhaus, and M.~S. Dresselhaus.
\newblock {\em Physical Properties of Carbon Nanotubes}.
\newblock Imperial College Press, 1998.

\bibitem{bostwick2007b}
A.~Bostwick, T~Ohta, J.~L. McChesney, T.~Seyller, K.~Horn, and E~Rotenberg.
\newblock Renormalization of graphene bands by many-body interactions.
\newblock {\em Solid State Communications}, 2007 (in press).

\bibitem{ohta2006}
T~Ohta, B~Bostwick, T.~Seyller, K.~Horn, and E.~Rotenberg.
\newblock Controlling the electronic structure of bilayer graphene.
\newblock {\em Science}, 313:951--954, 2006.

\bibitem{bostwick2007}
Aaron Bostwick, Taisuke Ohta, Thomas Seyller, Karsten Horn, and Eli Rotenberg.
\newblock Quasiparticle dynamics in graphene.
\newblock {\em Nat Phys}, 3(1):36--40, 2007.

\bibitem{wehling2007}
T.~O. Wehling, K.~S. Novoselov, S.~V. Morosov, E.~E. Vdovin, M.~I. Katsnelson,
  A.~K. Geim, and A.~I. Lichtenstein.
\newblock Molecular doping of graphene.
\newblock {\em cond-mat/0703390}, 2007.

\bibitem{damascelli2003}
A~Damascelli, Z.~Hussain, and Z.~X. Shen.
\newblock Angle-resolved photoemission studies of the cuprate superconductors.
\newblock {\em Review of Modern Physics}, 75:473, 2003.

\bibitem{kaminski2005}
Adam Kaminski and Helen~M. Fretwell.
\newblock On the extraction of the self-energy from angle-resolved
  photoemission spectroscopy.
\newblock {\em New Journal of Physics}, 7:98, 2005.
\newblock 1367-2630.

\bibitem{strocov2001}
V.~N. Strocov, A.~Charrier, J.~M. Themlin, M.~Rohlfing, R.~Claessen,
  N.~Barrett, J.~Avila, J.~Sanchez, and M.~C. Asensio.
\newblock Photoemission from graphite: Intrinsic and self-energy effects.
\newblock {\em Phys. Rev. B}, 64(7):075105, 2001.

\bibitem{kihlgren2002}
T.~Kihlgren, T.~Balasubramanian, L.~Walldén, and R.~Yakimova.
\newblock Narrow photoemission lines from graphite valence states.
\newblock {\em Physical Review B}, 66(23):235422, 2002.

\bibitem{soe2004}
W.~H. Soe, K.~H. Rieder, A.~M. Shikin, V.~Mozhaiskii, A.~Varykhalov, and
  O.~Rader.
\newblock Surface phonon and valence band dispersions in graphite overlayers
  formed by solid-state graphitization of 6h-sic(0001).
\newblock {\em Phys. Rev. B.}, 70(11):115421--6, 2004.

\bibitem{rollings2006}
E.~Rollings, G.~H. Gweon, S.~Y. Zhou, B.~S. Mun, J.~L. McChesney, B.~S.
  Hussain, A.~V. Fedorov, P.~N. First, W.~A. de~Heer, and A.~Lanzara.
\newblock Synthesis and characterization of atomically thin graphite films on a
  silicon carbide substrate.
\newblock {\em Journal of Physics and Chemistry of Solids},
  67(9-10):2172--2177, 2006.

\bibitem{seyller2006}
Th~Seyller, K.~V. Emtsev, F.~Speck, K.~Y. Gao, and L.~Ley.
\newblock Schottky barrier between 6h-sic and graphite: Implications for
  metal/sic contact formation.
\newblock {\em Applied Physics Letters}, 88(24):242103--3, 2006.

\bibitem{emtsev2006}
KV~Emtsev, T.~Seyller, F~Speck, L~Ley, P~Stojanov, JD~Riley, and RGC Leckey.
\newblock Initial stages of the graphite-sic(0001) interface formation studied
  by photoelectron spectroscopy.
\newblock {\em cond-mat}, page 0609383, 2006.

\bibitem{ohta2007}
Taisuke Ohta, Aaron Bostwick, J.~L. McChesney, Thomas Seyller, Karsten Horn,
  and Eli Rotenberg.
\newblock Interlayer interaction and electronic screening in multilayer
  graphene investigated with angle-resolved photoemission spectroscopy.
\newblock {\em Physical Review Letters}, 98(20):206802--4, 2007.

\bibitem{varchon2007}
F.~Varchon, R.~Feng, J.~Hass, X.~Li, B.~N. Nguyen, C.~Naud, P.~Mallet, J.-Y.
  Veuillen, C.~Berger, E.~H. Conrad, and L.~Magaud.
\newblock Electronic structure of epitaxial graphene layers on sic: effect of
  the substrate.
\newblock {\em cond-mat/}, page 0702311, 2007.

\bibitem{mattausch2007}
Alexander Mattausch and Oleg Pankratov.
\newblock Ab initio study of graphene on sic.
\newblock {\em cond-mat/07040216}, 2007.

\bibitem{mallet2007}
Pierre Mallet, Francois Varchon, Cecile Naud, Laurence Magaud, Claire Berger,
  and Jean-Yves Veuillen.
\newblock Electron states of mono- and bilayer graphene on sic probed by stm.
\newblock {\em cond-mat/}, page 0702406, 2007.

\bibitem{divincenzo1984}
D.~P. DiVincenzo and E.~J. Mele.
\newblock Self-consistent effective-mass theory for intralayer screening in
  graphite intercalation compounds.
\newblock {\em Phys. Rev. B}, 29:1685, 1984.

\bibitem{mccann2006}
E.~McCann and V.I. Fal'ko.
\newblock Landau-level degeneracy and quantum hall effect in a graphite
  bilayer.
\newblock {\em Phys. Rev. Lett.}, 96:086805, 2006.

\bibitem{zhou06}
S.~Y. Zhou, G.~H. Gweon, and A.~Lanzara.
\newblock Low energy excitations in graphite: the role of dimensionality and
  lattice defects.
\newblock {\em Annals of Physics}, 321:1730--46, 2006.

\bibitem{sugawara07}
K.~Sugawara, T.~Sato, S.~Souma, T.~Takahashi, and H.~Suematsu.
\newblock Anomalous quasiparticle lifetime and strong electron-phonon coupling
  in graphite.
\newblock {\em Phys. Rev. Lett.}, 98(3):036801--4, 2007.

\bibitem{lanzara2007}
A.~Lanzara.
\newblock First direct observation of dirac fermions in graphite.
\newblock {\em Bull. Amer. Phys. Society}, 2007.

\bibitem{zhou2007}
S.~Y. Zhou, G.~H. Gweon, J.~Graf, D.~Siegel, E.~Rollings, and A.~Lanzara.
\newblock Arpes study of the electronic dynamics from graphene to graphite.
\newblock {\em Bull. Amer. Phys. Society}, 2007.

\bibitem{rotenberg2003}
E.~Rotenberg, H.~Koh, K.~Rossnagel, H.W. Yeom, J.~SchŠfer, B.~Krenzer, M.P.
  Rocha, and S.D. Kevan.
\newblock In $\sqrt{7}\times\sqrt{3}$ on si(111): a nearly free electron metal
  in two dimensions.
\newblock {\em Phys. Rev. Lett.}, 91:246404, 2003.

\bibitem{crain2002}
J.~N. Crain, K.~N. Altmann, C.~Bromberger, and F.~J. Himpsel.
\newblock Fermi surfaces of surface states on si(111)-ag, au.
\newblock {\em Phys.Rev. B}, 66(20):205302, 2002.

\bibitem{shirley1995}
EL~Shirley, LJ~Terminello, A~Santoni, and F.~J. Himpsel.
\newblock Brillouin-zone-selection effects in graphite photoelectron angular
  distributions.
\newblock {\em Phys. Rev. B}, 51(19):13614--22, 1995.

\bibitem{kordyuk2005}
A.~A. Kordyuk, S.~V. Borisenko, A.~Koitzsch, J.~Fink, M.~Knupfer, and
  H.~Berger.
\newblock Bare electron dispersion from experiment: Self-consistent self-energy
  analysis of photoemission data.
\newblock {\em Phys. Rev. B}, 71(21):214513--11, 2005.

\bibitem{hwang2006b}
E.~H. Hwang, B.~Yu-Kuang~Hu, and S.~Das~Sarma.
\newblock Inelastic carrier lifetime in graphene.
\newblock {\em cond-mat}, page 0612345, 2006.

\bibitem{valla1999}
T.~Valla, A.V. Fedorov, P.D. Johnson, and S.L. Hulbert.
\newblock Many-body effects in angle-resolved photoemission: quasiparticle
  energy and lifetime of a mo(110) suface state.
\newblock {\em Phys. Rev. Lett.}, 83:2085--8, 1999.

\bibitem{hengsberger1999}
M.~Hengsberger, D.~Purdie, P.~Segovia, M.~Garnier, and Y.~Baer.
\newblock Photoemission study of a strongly coupled electron-phonon system.
\newblock {\em Phys. Rev. Lett.}, 83:592--5, 1999.

\bibitem{rotenberg2000}
E.~Rotenberg, J.~Schaefer, and S.D. Kevan.
\newblock Coupling between adsorbate vibrations and an electronic surface
  state.
\newblock {\em Phys. Rev. Lett.}, 84:2925--8, 2000.

\bibitem{calandra2005}
M.~Calandra and F.~Mauri.
\newblock Theoretical explanation of superconductivity in c6ca.
\newblock {\em Phys. Rev. Lett.}, 95:237002, 2005.

\bibitem{calandra2006}
M.~Calandra and F.~Mauri.
\newblock Origin of superconductivity of cac6 and of other intercalated
  graphites.
\newblock {\em physica status solidi (b)}, 243(13):3458--3463, 2006.

\bibitem{grimvall1981}
G.~Grimvall.
\newblock {\em The Electron Phonon Interaction in Metals}.
\newblock North Holland Publishing Company, 1981.

\bibitem{vafek2006}
Oskar Vafek.
\newblock Thermoplasma polariton within scaling theory of single-layer
  graphene.
\newblock {\em Phys. Rev. Lett.}, 97(26):266406--4, 2006.

\bibitem{stern1967}
Frank Stern.
\newblock Polarizability of a two-dimensional electron gas.
\newblock {\em Phys. Rev. Lett.}, 18(14):546, 1967.

\bibitem{kliewer1973}
K.~L. Kliewer and H.~Raether.
\newblock Plasmon observation using x rays.
\newblock {\em Phys. Rev. Lett.}, 30:971, 1973.

\bibitem{hodges1971}
Christopher Hodges, Henrik Smith, and J.~W. Wilkins.
\newblock Effect of fermi surface geometry on electron-electron scattering.
\newblock {\em Phys. Rev. B}, 4(2):302, 1971.

\bibitem{mccann2006b}
E.~McCann.
\newblock Asymmetry gap in the electronic band structure of bilayer graphene.
\newblock {\em cond-mat}, page 0608221, 2006.

\bibitem{guinea2006}
F.~Guinea, A.~H. Castro~Neto, and N~M~R Peres.
\newblock Electronic states and landau levels in graphene stacks.
\newblock {\em cond-mat}, page 0604396, 2006.

\bibitem{min2007}
Hongki Min, Bhagawan Sahu, Sanjay~K. Banerjee, and A.~H. MacDonald.
\newblock Ab initio theory of gate induced gaps in graphene bilayers.
\newblock {\em Phys. Rev. B}, 75(15):155115--7, 2007.

\bibitem{wang2006b}
Z.~F. Wang, Qunxiang Li, Haibin Su, Xiaoping Wang, Q.~W. Shi, Jie Chen, Jinlong
  Yang, and J.~G. Hou.
\newblock Electronic structure of bilayer graphene: A real-space green's
  function study.
\newblock {\em Phys. Rev. B.}, 75(8):085424--8, 2007.

\bibitem{aoki2007}
M.~Aoki and H.~Amawashi.
\newblock Dependence of band structures on stacking and field in layered
  graphene.
\newblock {\em cond-mat/0702257}, 2007.

\bibitem{nilsson2006}
J.~Nilsson, A.~H. Castro~Neto, F.~Guinea, and N.~M.~R. Peres.
\newblock Transmission through a biased graphene bilayer barrier.
\newblock {\em cond-mat/0607343}, 2006.

\bibitem{castro2006}
E.~V. Castro, K.~S. Novoselov, S.~V. Morosov, N.~M.~R. Peres, J.~M.~P.
  Lopes~dos Santos, J.~Nilsson, F.~Guinea, A.~K. Geim, and A.~H. Castro~Neto.
\newblock Biased bilayer graphene: semiconductor with a gap tunable by electric
  field effect.
\newblock {\em cond-mat/0611342}, 2006.

\bibitem{pivetta2005}
Marina Pivetta, Francois Patthey, Ingo Barke, Heinz Hovel, Bernard Delley, and
  Wolf-Dieter Schneider.
\newblock Gap opening in the surface electronic structure of graphite induced
  by adsorption of alkali atoms: Photoemission experiments and density
  functional calculations.
\newblock {\em Phys. Rev. B}, 71(16):165430--4, 2005.

\end{thebibliography}

\end{document}